%% file: ms.tex
\documentclass[12pt,preprint]{aastex}

\usepackage{multirow}

\shorttitle{Orion NIRSPEC}
\shortauthors{}

\begin{document}

\title{The Spectroscopically Determined Substellar
Mass Function of the Orion Nebula Cluster}

\author{Catherine L. Slesnick\altaffilmark{1}, Lynne
A. Hillenbrand\altaffilmark{1}, \& John M. Carpenter\altaffilmark{1}}

\affil{Dept.\ of Astronomy, MS105-24, California Institute of
Technology,Pasadena, CA 91125}

\email{cls@astro.caltech.edu, lah@astro.caltech.edu, jmc@astro.caltech.edu}

\altaffiltext{1}{Visiting astronomer, W. M. Keck Observatory, which
is operated as a scientific partnership among the California Institute of Technology, the university of California and the National Aeronautics and Space Administration.}

\begin{abstract}

We present a 
spectroscopic study of candidate  
brown dwarf members of the Orion Nebula Cluster (ONC).  We obtained new $J$-- 
and/or $K$--band  
spectra of $\sim$100 objects within the ONC which are expected to be substellar
based on their $K,(H-K)$ magnitudes and colors.
Spectral classification in the near-infrared of young low mass objects
is described, including the effects of surface gravity, veiling due to
circumstellar material, and reddening.
From our derived spectral types and existing near-infrared photometry
we construct an HR diagram for the 
cluster.
Masses are inferred for each object and used to derive the brown dwarf fraction and
assess the  mass function for the inner 5.'1 $\times$ 5.'1 of the ONC,
down to $\sim$0.02 M$_\odot$.  The logarithmic mass function rises to a peak at 
$\sim$0.2 M$_\odot$,
similar to previous IMF determinations derived from purely photometric methods, but falls off 
more sharply at the hydrogen-burning limit before
leveling through the substellar regime.
We compare the mass function
derived here for the inner ONC to those 
presented in recent literature for the sparsely populated
Taurus cloud members and the rich cluster IC 348.
We 
find good agreement between the shapes and peak values of the ONC
and IC 348 mass distributions, but little similarity between
the ONC and Taurus results. 

\end{abstract}


\section{Introduction}

The stellar mass and age distributions in young star clusters
can help answer some of the fundamental
questions of cluster formation theory: Do all cluster members 
form in a single burst 
or is star formation a lengthy process?  
Is the distribution of stellar 
masses formed during a single epoch within a cluster, 
also known as the initial mass function (IMF), universal or does it vary
with either star formation environment or time?
While the stellar mass function 
has long been studied (e.g., Salpeter 1955), we are only 
recently beginning to explore the very low mass end of the distribution 
into the substellar regime.  
Identification of large, unbiased samples of low mass objects, especially 
in star-forming regions, is crucial to our understanding of the formation and early evolution of low mass
stars and brown dwarfs. 

Young stellar clusters are particularly valuable for examining the 
shape of the low mass IMF because 
the lowest mass members have not yet been lost to dynamical evolution.
Furthermore, contracting low-mass pre-main sequence stars and brown dwarfs 
are 2-3.5 orders of magnitude more luminous than their counterparts on the main sequence, 
and thus can be more readily detected in large numbers.
The dense molecular clouds associated with star-forming regions also
reduce background field star contamination.


Because the ONC is one of the nearest massive star-forming regions to the Sun and the 
most populous young cluster within $\sim$2 kpc, it has been observed at virtually all wavelengths
over the past several decades.  
However, only recently have increased sensitivities due to near-IR detectors on
larger telescopes allowed us to
begin to understand and characterize the extent of the ONC's young stellar and brown dwarf
population which, at $\lesssim$ 1-2 Myr, is just beginning to emerge from its giant molecular cloud. 

Several recent studies have explored the ONC at substellar masses.  
Hillenbrand \& Carpenter (2000) (hereafter HC00) present the results of an $H$ and $K$  
imaging survey of the inner ~5'.1 x 5'.1 region of the ONC.
Observed magnitudes, colors, and star counts were used to constrain the shape of the ONC mass function across the hydrogen burning limit down to $\sim$0.03 M$_\odot$.  They find evidence in the 
log-log mass 
function for a turnover above the hydrogen-burning
limit, then a plateau 
into the substellar regime.  
A similar
study by Muench et al. (2002; hereafter M02) uses $J,H,K$ imaging of the ONC 
to derive an IMF which rises to a broad primary peak at the lowest stellar masses between 0.3 M$_\odot$ and the hydrogen 
burning limit before turning over and declining into the substellar regime.  However, instead of a plateau through the lowest masses, M02 find evidence for a secondary peak between 0.03--0.02 M$_\odot$.
Luhman et al. (2000)
use $H$ and $K$ infrared imaging and limited ground-based spectroscopy to constrain the mass function and
again find a peak just above the substellar regime, but then a steady decline through the lowest mass objects.

Generally speaking, $J,H,K$ photometry alone is insufficient
for deriving stellar/substellar masses, though may be adequate in a statistical 
sense for estimating mass distributions given the right assumptions.
The position of a young star in a near-IR color-magnitude diagram (CMD)
is dependent on mass, age, extinction, and the possible presence 
of a circumstellar disk.  
These characteristics
affect the conversion of a star's infrared magnitude and color into 
its stellar mass.  Unless the 
distributions of these parameters are known a priori, knowledge of the cluster's
luminosity function alone is not sufficient to draw definitive conclusions about its mass function.
In addition, cluster membership is often poorly known
and statistical estimates concerning the extent and characterization
of the field star population must be derived.  In the case of the densely populated
ONC, it has been suggested
that the field star contamination 
is small but non-negligible toward fainter magnitudes.  
HC00 used a modified version of the
Galactic star count model (Wainscoat et al. 1992) convolved with a local extinction map
(derived from a C$^{18}$O molecular line map) to estimate the field star contribution, which they found to constitute $\sim$5\% of the stars down to 
their completeness limit at $K\sim$17.5.

In order to study a cluster's IMF in more than just a statistical sense, 
spectroscopy is needed to confirm cluster membership of  
individual stars and uniquely determine
location in the HR diagram (and hence mass). 
We have obtained near-infrared spectra of 
97
stars in the ONC.
This wavelength regime (1-2.5 $\mu$m) is of extreme interest for very cool stars and brown dwarfs not only because ultracool objects emit the bulk of their energy in the near-infrared, but also because this regime contains temperature-sensitive atomic as well as broad molecular features. In addition, there are several diagnostic lines which can be used as surface gravity indicators.  
From analysis of these data combined with existing photometry and pre-main sequence
evolutionary theory we construct the cluster's IMF across the 
substellar boundary.  We then compare our results to those found from previous studies,
both of the ONC and of clusters similar in age to the ONC but which have 
different star-forming environments. 

In Section 2 we describe our data acquisition and reduction.  In Section 3 we 
present our spectra and methods for spectral classification.  This section includes 
a discussion of the effects of extinction and veiling due to circumstellar disks.
In Section 4 we create an HR diagram for the stellar and substellar objects for which we have new 
spectral types. 
Section 5 contains our derivation of the ONC's IMF and Section 6 our analysis
and comparison to previous work.

\section{Observations}

\subsection{Infrared Spectroscopic Sample Selection}

\begin{figure}
\epsscale{0.8}
\plotone{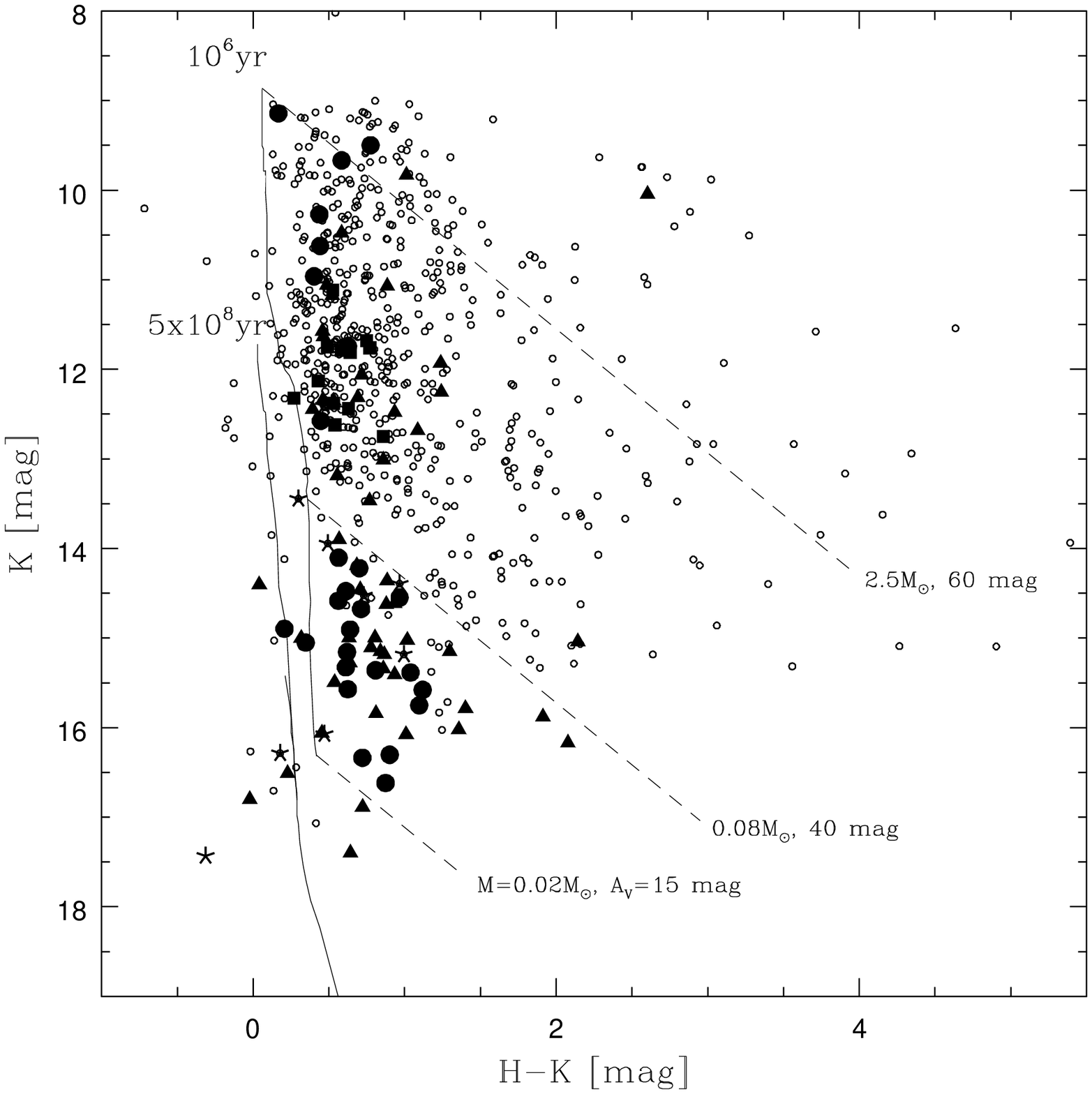}
\caption{$K$, ($H-K$) color-magnitude diagram for the ONC.  
Isochrones are from DM97 
and have been 
transformed into the $K$, ($H-K$) plane
assuming the age ($\lesssim$ 1 Myr) and distance (480 pc) of the ONC.
Dashed lines are reddening vectors emanating at the indicated masses from the 1 Myr isochrone.
Open circles
represent HC00 photometry of the inner 5'.1 $\times$ 5'.1 of the nebula.  
Sources for which we have $J$-band or $K$-band spectra taken with NIRSPEC are marked as solid circles
and triangles, respectively. 
Stars were selected for spectroscopy based on their
location on the CMD below 
0.08 $M_\odot$. 
Often it was possible to place multiple stars on the slit due to the high stellar density of the cluster, allowing us to observe several 
brighter (more massive) stars.
Starred points indicate sources for which we have new red optical LRIS spectral
classifications.
Solid squares represent sources (mostly in the outer nebula) for which we have $J$ \& $K$-band spectra taken 
with CRSP.  
}
\label{fig:cmd}
\end{figure}

Candidate sub-stellar objects have been selected from 
the HC00 photometric survey   
for follow-up spectroscopy 
to determine if their temperatures and surface gravities are consistent with those of brown dwarf objects at the age ($\sim$ 1 Myr) and distance (480 pc)
of the ONC.  
Figure~\ref{fig:cmd} is a color-magnitude diagram showing the HC00 photometry.
We transformed the isochrones and mass tracks of D'Antona \& Mazzitella 1997 \& 1998\footnote{The 1998 models are a web-only correction at
$<$0.2 M$_\odot$ to their original 1997 work.} (hereafter DM97)
into the $K,(H-K)$ plane and selected stars for spectroscopy based on their
location in the CMD below a line which corresponds to the reddening vector 
originating at the hydrogen burning limit (M = 0.08 M$_\odot$). 
In addition to stars selected from infrared photometry, several stars 
 were 
followed-up with infrared 
spectroscopy before the deep HC00 data became available 
based on optical ($V,I$) magnitudes and colors (Hillenbrand 1997; hereafter H97)
indicating they might be substellar.  

We obtained new infrared spectra of 97 objects within the ONC; 81 selected from the HC00 work were observed using NIRSPEC (Section 2.2.1)
and 16 selected from optical magnitudes and colors were observed 
using CRSP (Section 2.2.2).  
Our sample included
$\sim$50\% of the stars in the HC00 survey area expected to be brown dwarfs based on their $K,(H-K)$ magnitudes 
and colors, down to the completeness limit 
of $K\sim$17.5.  
Often it was possible to place multiple stars on the slit due to the high stellar density of the cluster, allowing us to observe several brighter stars, some with known spectral types
which could be used as secondary spectral standards.

\subsection{Infrared Data Acquisition}

\subsubsection{NIRSPEC Data}

Near-IR spectra of candidates selected from infrared photometry 
were taken with the NIRSPEC spectrograph 
on the Keck II Telescope on Mauna Kea,
Hawaii.  This near-infrared spectrograph has a 1024x1024 InSb array with 27 $\mu$m pixels.  
We used the low-resolution mode of the camera with an entrance slit of 42''x0.72'', 
resulting in a resolving power of R$\sim$1400.  Stars were placed in the entrance slit using a $K$-band guide camera.
$J$-band spectra were taken on 2 \& 3 February, 2002 
through the NIRSPEC-3 filter which spans the wavelength 
range of 1.143--1.375 $\mu$m.  Typical exposure times were $\sim$300 s.
$K$-band data were taken on 29 November, and 1 \& 2 December, 2002, 
through the K' filter (1.950--2.295 $\mu$m).
This filter offers less wavelength coverage than either the K or the NIRSPEC-7 filters;
however it afforded us extremely high transmission, particularly at the 
short wavelength end.  
The K' filter
has $>$ 90\% transmission from 2--2.95$\mu$m, 
whereas the NIRSPEC-7 filter gives an average of $\sim$80\% transmission (sometimes as low as 70\%) in this range
and the K filter gives an average of 70\%.  Our aim when taking the $K$-band
spectra was to observe as many faint objects as possible so that
we would have a statistically large enough sample of 
known cluster members with spectral types
(and therefore, masses) from which to construct an accurate IMF.  
We felt it would be a more effective use of limited telescope time to 
use the K' filter 
so that we could observe more and fainter objects
with shorter exposure times.  We are still able to make use of 2 of the 3 
molecular absorption features between 2--2.5 $\mu$m commonly utilized for classification
of low mass stars (see Section $3.1.2$).   
Typical exposure times for the $K$-band data were 300--600 s.

For each object, we took 
exposures in sets of 3, nodding the star along the slit between each exposure.  For our fainter targets,
2--3 sets of observations were required.  In most cases
we were able to observe multiple objects in a single exposure by rotating the slit 
position angle.  
Telluric reference stars were observed throughout the night over a wide range in airmass.
We observed O \& G stars whose spectra have relatively few absorption features
in the $J$ \& $K$-bands at our resolution and may be used as telluric templates.  
NeAr arc lamp spectra and internal flats used in rectifying the images
(see Section 2.3) were taken at the start and end 
of each night. 

\subsubsection{CRSP Data}

For sources selected from optical $I,(V-I)$ magnitudes and colors,
spectra were obtained with the Cryogenic Infrared Spectrograph (CRSP) 
on the Mayall 4m telescope at Kitt Peak National Observatory on the 
nights of 16--19 January, 1998.  
The instrument is a cooled grating spectrograph with a 256x256 InSb 
detector.  Low resolution (R = 1000) spectra were collected using a 200 l/mm
grating in the Y--, J--, H--, and K--bands (3rd, 2nd, 2nd, and 1st orders
respectively).  The slit subtended  1.0'' with the infrared
seeing typically 0.9--1.2''.  Stars were positioned in the slit via a visible 
guide camera and data were collected in a standard two-beam beam--switching 
mode.  Integration times ranged from 0.316 (minimum read-out time) 
to 120 seconds for individual frames.  
A0 and G2 telluric standard stars were observed for
correction of spectral standards while the O7 star $\theta^1$C Ori was
used for program stars.  Wavelength calibration was established from exposures
of a 
HeNeAr arc lamp while dome--flats were collected with and
without incandescent illumination of a white screen mounted inside the
telescope dome. 

\subsection{Infrared Data Reduction}

\subsubsection{NIRSPEC Data}

We reduced the NIRSPEC data within IRAF, applying both standard tasks 
and custom techniques developed for reducing NIRSPEC data.  All sources, including standards,
were pre-processed and extracted in the same manner as outlined below.
 
Bad pixel masks were created by median-combining dark exposures taken at the 
beginning of each night.  
The mask was then applied to all images within a given night using the IRAF FLVMASK task which 
replaces bad pixels by interpolating
along the spectral direction.  Cosmic rays were removed using QZAP, 
written by Mark Dickenson.   
Since raw NIRSPEC data are tilted in both the spatial and dispersion directions, 
we used custom IRAF software written by Gregory D. 
Wirth (available publicly on the NIRSPEC website) to rectify the data.  
This software first removes
shifts in the x (spatial) direction using a spatial map, created from a flat-field 
exposure, to align the edge of an image along a single column.  
Next, arc lamp spectra are used to align emission features along image rows
in the wavelength direction
by applying a second-order chebyshev fit.  

We did not flat field the data because we found the internal flats did not 
uniformly illuminate the detector in the lower resolution mode.  In addition, for the $J$-band data presence of 
ice on the dewar window caused 
time-variable streaks on the flat field images in the spectral direction.  
While non-uniform illumination of the flat field images
 does not cause a large problem for observations concerned only with narrow  
absorption/emission lines, we are interested in the strength of broad molecular absorption
bands measured as relative flux levels at 
different parts of the spectrum.  Therefore, it is essential that the data be flat 
in the dispersion direction.  We found using the flat-field  
images on the standard star observations introduced
error rather than correcting it.  Instead, division by a telluric reference 
star corrected adequately for any 
 non-uniform features in the dispersion direction of the detector.
Once the effectiveness of this method was confirmed with our sequences of
standard stars, we applied it to our ONC spectra (see below).
 

To remove background night sky and nebular emission lines, we used one of the adjacent nod position images to create 
a sky frame (or an average of two nod position images).  Simple image subtraction does not work for this data because
the background emission varies on short timescales. 
Instead, we used the median value along each row to create background emission frames free from stellar continuum
for both the target and sky frames.  The sky emission frame was then scaled at each row to match the relative intensities 
of the emission lines on the target emission frame.  This new sky frame was then subtracted off the original target frame.  
Because our objects
are located in the Orion Nebula whose spectrum
contains very strong emission features 
which can vary on extremely small spatial scales (a few arcseconds), some residual emission/absorption areas were sometimes left after sky subtraction. 
To correct for these, individual background regions were 
determined and subtracted from the source during spectral 
extraction (using APALL
within IRAF).
 
Each spectrum was wavelength calibrated using sky emission lines, except the 
short exposures for which the next closest observation in time was used.
A telluric spectrum was produced for each extraction from a weighted average of the two flux standards closest in airmass, thereby creating a standard spectrum at the airmass of the object.
For most objects we used 
$\theta^1$C Ori as the standard star
both because of its near-featureless spectrum, and because
its location near the center of the Orion Nebula 
made it easy to observe at representative airmasses.   
The only exceptions were some of the M-type standard stars, particularly those in 
Upper Sco, which were observed at airmasses $\geq$ 1.6.  
For these objects we used HR4498 (G0 V) as a telluric standard after interpolating over the few
detectable absorption features in its spectrum.  
Finally, we averaged together all spectra for a given object 
and multiplied by a synthetic blackbody spectrum
appropriate for the temperature of the telluric standard used.  
The spectra were normalized to have an average flux of unity (over the entire wavelength region within
the filter) for analysis.

\begin{figure}
\epsscale{0.8}
\plotone{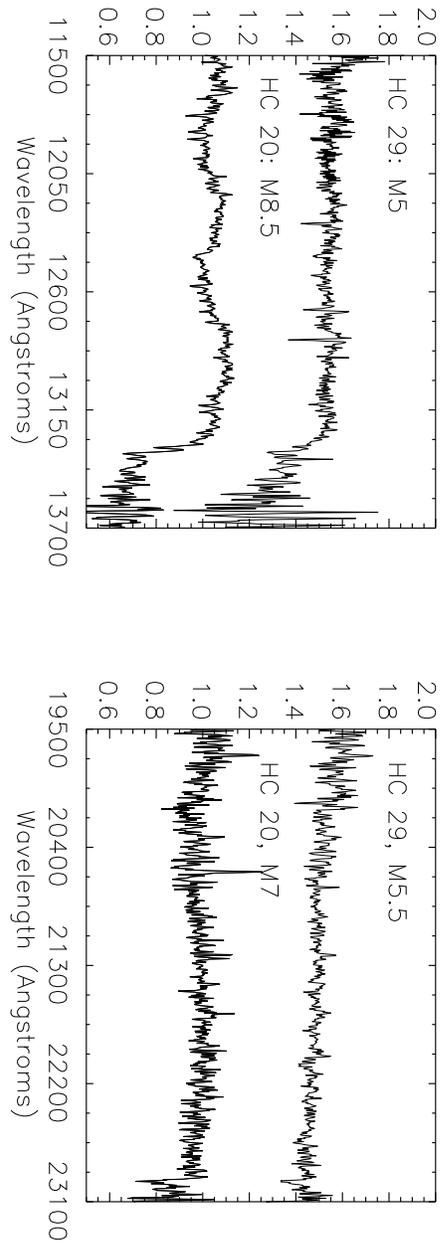}
\caption{
NIRSPEC spectra representative of the typical S/N for our program objects.  Both targets have relatively low extinction (A$_V$ $<$ 5).
Derived spectral types are indicated and demonstrate the degree of agreement
between $J$ and $K$-band spectral types when both 
data are available.}
\label{fig:orion}
\end{figure}

Figure~\ref{fig:orion} shows two NIRSPEC spectra representative of the typical S/N for our program objects.  We have chosen to show these objects
because they were  
observed during both observing runs (along with HC 383, HC 509 and several of the standards)
 as a consistency check.
In addition, they have relatively low extinction (A$_V$ $<$ 5) and therefore
can be more easily compared to previously published spectra of objects 
at the same temperatures (see Section 3.4).
The few residual emission lines from the background nebula have been interpolated over.
In general, we find agreement to within errors between $J$ \& $K$-band spectral types when both 
data are available.
A detailed description of our spectral classification methods, including a discussion
of reddening estimates is given in Section 3.  Typical signal-to-noise
ratios of the ONC spectra were $\sim$10-70 pixel$^{-1}$, with nearly all spectra classifiable.


\subsubsection{CRSP Data}

Standard image processing was accomplished using IRAF. After linearization,
image trimming, and bad pixel correction, 
individual beam--switched pairs of spectra were subtracted in order to 
effect a first--order sky subtraction and to remove dark current and bias
offsets.  Normalized flat--fields were constructed from 
median--filtered combinations of ``lights--on'' minus ``lights--off''
dome flats, and used to flat--field the data frames.
Spectra
were extracted using the APALL task and combined using a flux-weighted
average.  The dispersion solution was derived from extracted arc spectra
and individually adjusted using cross-correlation techniques to bring 
the spectra to a common wavelength zero point.  This step was necessary due 
to the continual shifts of the grating required in order to obtain data 
in all four near-infrared atmospheric windows.
To remove atmospheric and instrumentation effects from the data, 
the combined spectra were divided by telluric standards in which Paschen
and Brackett series lines were first interpolated over.  In order to 
preserve the intrinsic spectral energy distributions of the program 
stars the telluric--divided spectra were multiplied by a 
black--body function appropriate to the adopted telluric standard.  
The final spectra were normalized to have an average flux of unity over the entire 
spectral range. 
For the analysis presented herein we make use of only the J-band and K-band 
spectra from this data set.


\subsection{Optical Spectroscopy}

We have compared the spectral types derived from our new infrared spectra 
to those obtained from optical spectra, either as presented in the
tables of H97 or as newly updated by us based on data obtained
with Keck II / LRIS.  Data were obtained with LRIS on six different occasions 
between January 1998 and January 2002 by us and by additional observers including 
N. Reid and B. Schaefer.  Both single slit and multislit modes 
were employed.  Stars for the single slit observations were selected in a 
manner similar to that employed for the CRSP observations described above, 
namely based on location in the $I,(V-I)$ color-magnitude diagram.  The multi-slit 
observations were focused similarly to the NIRSPEC observations described
above, primarily on the brighter substellar candidates from the $K,(H-K)$ survey
of HC00.  Objects for which we have classifiable LRIS spectra are shown as starred points on Figure~\ref{fig:cmd}.

The grating was either a 400/8500 or 600/7500 (lines per mm / blaze)
depending on the run, 
producing R$\sim$1400 spectra nominally over 6000-10500 \AA$\;$ with shifts 
in spectral coverage of up to $\pm$1000\AA $\;$for the multislit data.
Single slit observations were typically 240-600 second integrations 
whereas the multi-slit data were taken in stacks of 5-8 600 second 
integrations in order
to avoid nebular saturation perpendicular to the dispersion direction and
thus across the slitlets.  Data were reduced, after median filtering and
stacking the multislit data, using standard techniques of bias-subtraction, 
flat-fielding, and spectral extraction with particular attention needed due 
to the strong and spatially variable nebular emission.
Spectra were classified using a depth of feature analysis as described 
in H97 with the TiO and VO bands the most important spectral
diagnostics in the M-type range of interest in this paper.

Optical spectral types are presented in Tables 1a and 1b for comparison to our
infrared spectral types.  Types which are not footnoted come from the new LRIS spectra
whereas footnoted types are from H97.

\section{Infrared Spectral Classification}

To ensure that we could accurately classify our program spectra 
we took a range of spectral main-sequence standards
using NIRSPEC: M0-M9 at $J$-Band
and K7-L3 at $K$-Band.  We also took spectra of mid to late M lower surface gravity 
stars in Upper Scorpius and Praesepe as well as field giants. 
For the $J$-band data 
we were able to include 
electronically available standard
spectra of nearby field dwarfs taken by the Leggett group\footnote{See www.jach.hawaii.edu/$\sim$skl/publications.html} because  
our classification indices rely
on the depth of H$_2$O and FeH absorption features, rather than the shape of the continuum (the continuum shape 
was later considered in comparison to our NIRSPEC standards when possible; see Section 3.1).
We did not have this luxury in the $K$-band due to the 
flat-field issues discussed in Section 2.3.1.  


As detailed below, we rely primarily on broad-band molecular features for 
spectral classification.
We present our spectral types for objects within the inner 5.'1$\times$5.'1
 of the ONC (centered on $\theta ^1$C) in Table 1a.  When possible, objects have been
cross-referenced to previous literature and optical spectral types are given. 
Table 1a also includes 
estimates for the surface gravity of each object based on the presence 
and relative strengths of atomic absorption lines.
Most (but not all -- see Sections 3.1 \& 3.2) 
of the atomic lines present in main sequence stars at similar temperatures are much weaker in the ONC spectra  
due to the young age and consequently lower surface gravity of the star.
This effect is enhanced if there is excess continuum 
emission from
a circumstellar disk. 
Therefore,
we do not use the properties of these lines to classify a star except to say, if 
the lines are present and strong, the star is likely not an ONC 
member.
Our procedures are discussed in Sections 3.1.1 \& 3.1.2
and result in FeH, H$_2$O-1 \& H$_2$O-2 index measurements also listed in Table 1a.

\subsection{$J$-Band Indices}
\begin{figure}
\epsscale{0.8}
\plotone{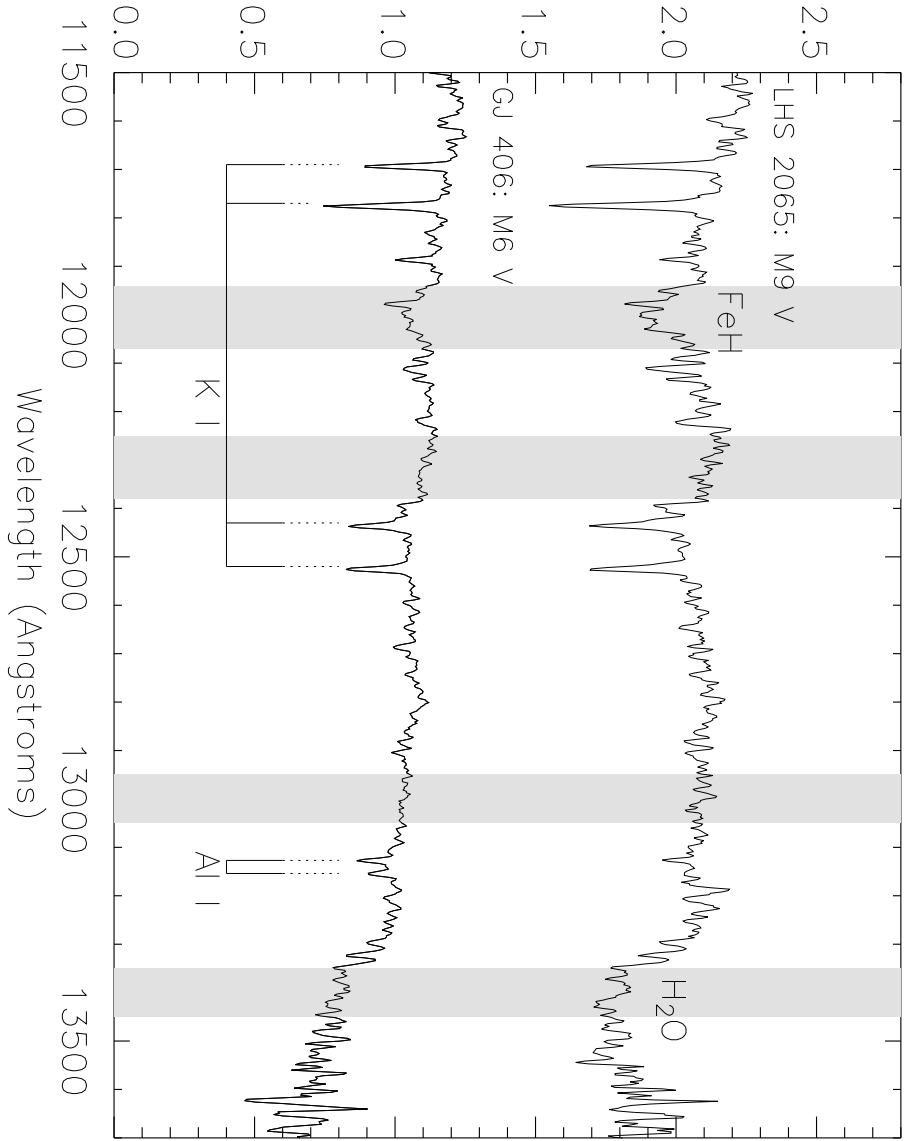}
\caption{$J$-band NIRSPEC spectra of standard stars 
at two temperatures.  The more 
prominent atomic (pairs of KI doublets and an Al I doublet) 
and temperature-sensitive molecular features (FeH and H$_2$O) are labeled.  Flux bands centered on molecular
and continuum features used in classification are shown as shaded regions.}
\label{fig:jtemp}
\end{figure}

\begin{figure}
\epsscale{0.8}
\plotone{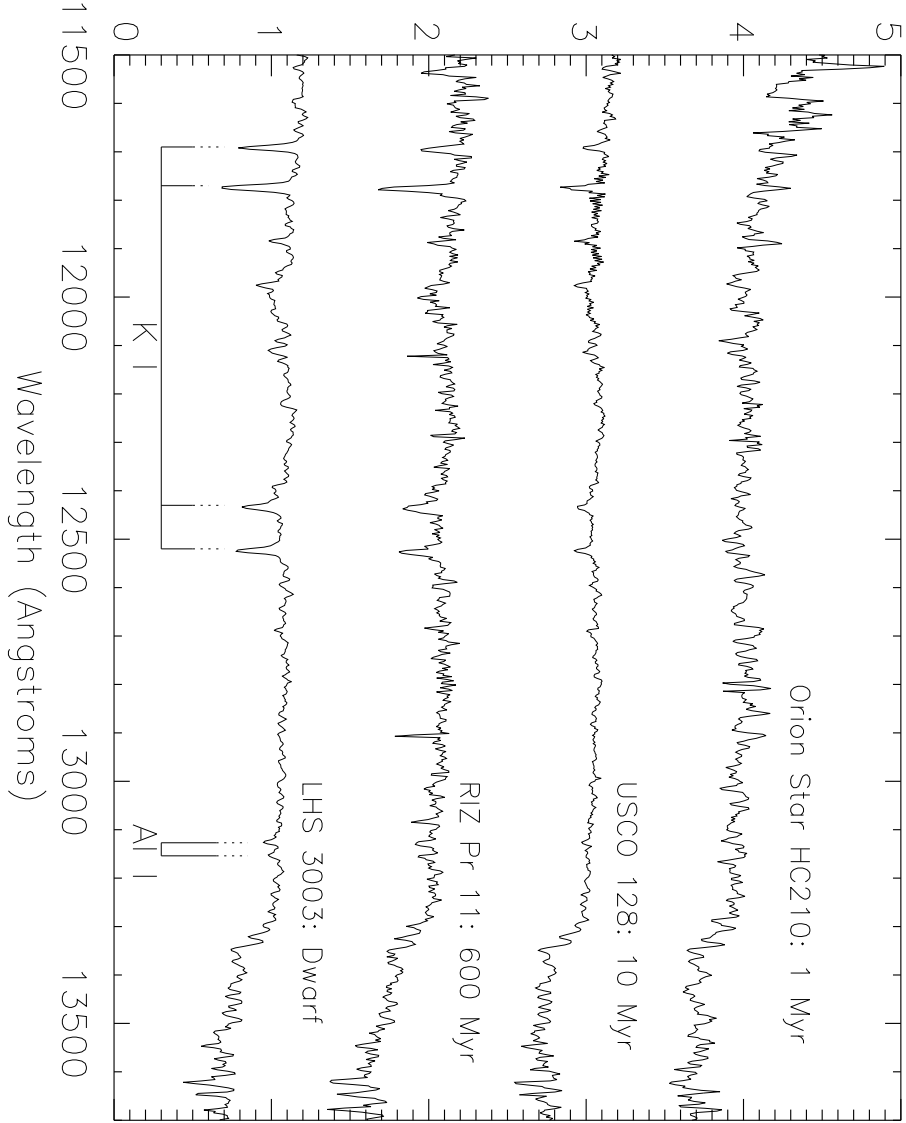}
\caption{$J$-band NIRSPEC data of four spectral class M7-M8 stars: an
optically-classified main sequence dwarf star (LHS 3003), a newly classified lower surface gravity star in Praesepe (RIZ Pr 11), an optically-classified star 
in Upper Sco
(USCO 128), and a newly classified Orion Star (HC210).  Strongly surface gravity sensitive
atomic features KI \& AlI are labeled.}
\label{fig:jgrav}
\end{figure}

Figure~\ref{fig:jtemp} shows two $J$-band M dwarf standard spectra at different temperatures, taken with NIRSPEC.  
The strongest atomic line transitions in this wavelength regime are the pairs of K I lines at 1.169 $\mu$m, 1.177 $\mu$m and 1.243 $\mu$m, 1.252 $\mu$m.
As can be seen, the strength of these features increases with later spectral type.
This trend is illustrated for a wider range of spectral types in Leggett et al. (1996) (stars with spectral types earlier than M7)
and McLean et al. (2003) (dwarfs of spectral type M6 and later).
However, the depth of the lines is also highly
dependent on surface gravity.  
Figure~\ref{fig:jgrav} shows four stars of spectral class M7--M8: an optically-classified dwarf star (LHS 3003), a newly classified lower surface gravity star of $\sim$600 Myr in Praesepe (RIZ Pr 11), an optically-classified star of $\sim$10 Myr in Upper Sco
(USCO 128), and a newly classified (see below) Orion Star of $\sim$1 Myr (HC210).  This figure illustrates the dramatic decrease in strength of the K I lines with decreasing surface gravity (see also McGovern et al. 2004 and Gorlova et al. 2003).  
We find these lines to be weak or absent in most of the ONC spectra, confirming cluster membership.  

The dominant molecular absorption features are iron hydride (FeH) and water (H$_2$O), the strongest bands of which are found at approximately 1.20 $\mu$m and  1.34 $\mu$m, respectively. 
Following the work of McLean et al (2000) and Reid et al (2001), we have constructed an index, H$_2$O-1, to measure the strength of the 1.34 $\mu$m water absorption feature using the ratio of the flux at 1.34 $\mu$m to that at 1.30 $\mu$m.  The positions of the flux bands have been shifted slightly from
those of previous authors 
to avoid 
contamination by variable emission features which arise in the surrounding nebula and may 
still be present after sky subtraction in some of the spectra.
In addition, we constructed a new index to measure the strength of the FeH feature using the ratio of flux at 1.20 $\mu$m to that at 1.23 $\mu$m.  Band widths are 100 \AA $\;$for the H$_2$O-1 index and 130 \AA $\;$for the FeH index.  
The feature and continuum bands are indicated by shaded regions on Figure~\ref{fig:jtemp}.   

\begin{figure}
\epsscale{0.8}
\plotone{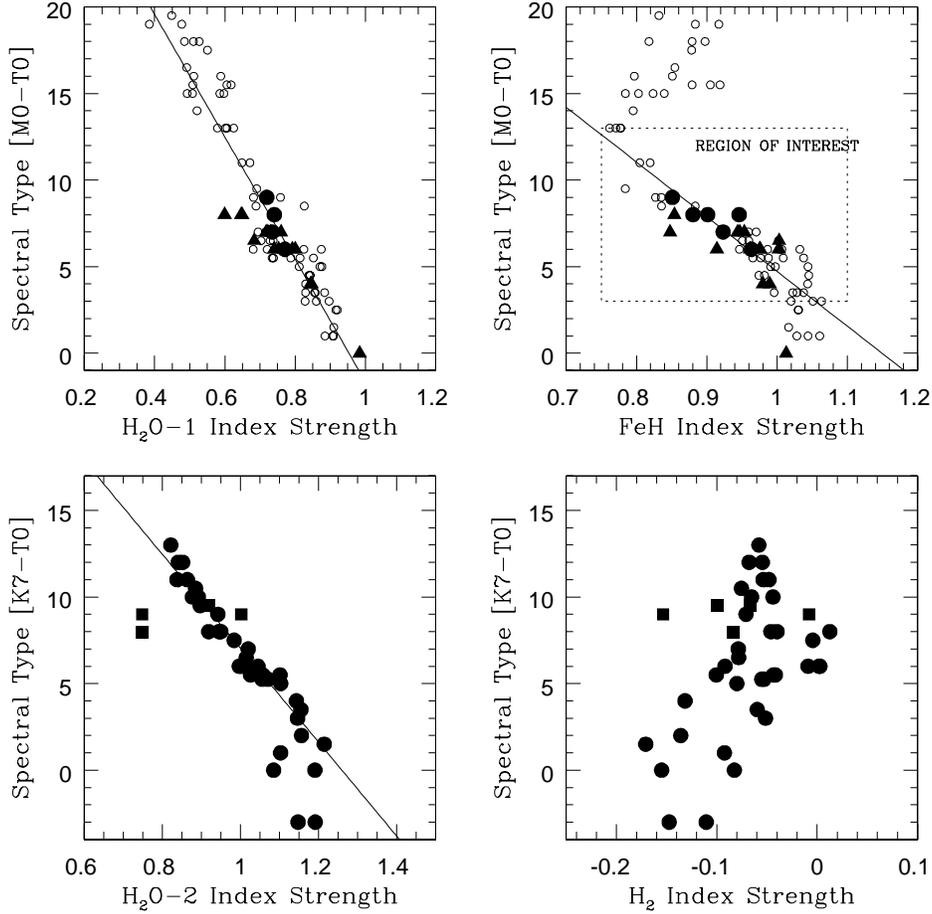}
\caption{Spectral type as a function of classification indices.  
Solid circles and triangles represent 
main-sequence standard stars, and lower surface 
gravity pre-main-sequence taken with NIRSPEC, respectively.  Solid squares correspond to giant stars.   
Open circles are nearby field
star and brown dwarf data from the Leggett group.
Solid lines are empirical fits listed in Table 2.
The FeH index works
 well only for a limited range in spectral type (M3-L3) where the absorption feature peaks in strength.
The H$_2$O-2 index works well for
objects later than M2.  We do not derive an empirical fit to the H$_2$ index because its dynamic range is small and the scatter significant.}
\label{fig:jh2o}
\end{figure}

We found the water index to be an excellent indicator of spectral type.  
The upper left panel of Figure~\ref{fig:jh2o} shows spectral type vs. the H$_2$O-1 index.  Here, a spectral type of 0 represents an M0 star and a spectral type of 10 represents an L0 dwarf.  In all panels, filled circles \& triangles represent standard star spectra taken with NIRSPEC at the same time as 
our program objects.
Open circles are the lower resolution data (R $\sim$400) taken from the Leggett group.  Despite the differences in equipment and data reduction techniques, we find excellent correlation between the two sets of data.   The strength of the water index decreases systematically with decreasing spectral type.  We derived a functional fit to the combined standard observations which we used to classify the Orion data.  Typical errors are $\sim$1.5 spectral subtypes.  All index fits are given in Table 2.

The upper right panel of Figure~\ref{fig:jh2o} shows spectral type vs. the FeH index.  As can be seen, this index works
 well only for a limited range in spectral type (M3--L3) where the absorption feature peaks in strength.   
However, 
stars earlier than M3 and later than L3 are classifiable by measuring the 
H$_2$O-1 index.
Furthermore, it is readily apparent via visual inspection of the spectra 
when a star has been classified by the FeH index as 
an M9 but is in fact an L6.  
Within the spectral range of interest, the strength of the FeH index decreases systematically with spectral type, similar to the H$_2$O-1 index.  We again derived a functional fit to the combined standard observations to be used in classifying our program objects (Table 2).  

In both top panels of Figure~\ref{fig:jh2o} the filled triangles, which represent 
the lower surface gravity standard stars, correspond to objects ranging in age from 
$\sim$600 Myr (Praesepe) to $\sim$10 Myr (Upper Sco \& TW Hydrae).  As can be seen, we find 
both molecular temperature (spectral type) indices to be stable to variations in surface gravity.
However, great care had to be taken in applying these indices because they are sensitive to effects from 
reddening and IR excess (see Sections 3.4 \& 3.5).  
Once preliminary 
spectral types were derived from both the H$_2$O-1 and FeH indices, 
all objects were 
inspected visually and compared to standards for confirmation or minor adjustment.  


\subsection{$K$-Band Indices}

\begin{figure}
\epsscale{0.8}
\plotone{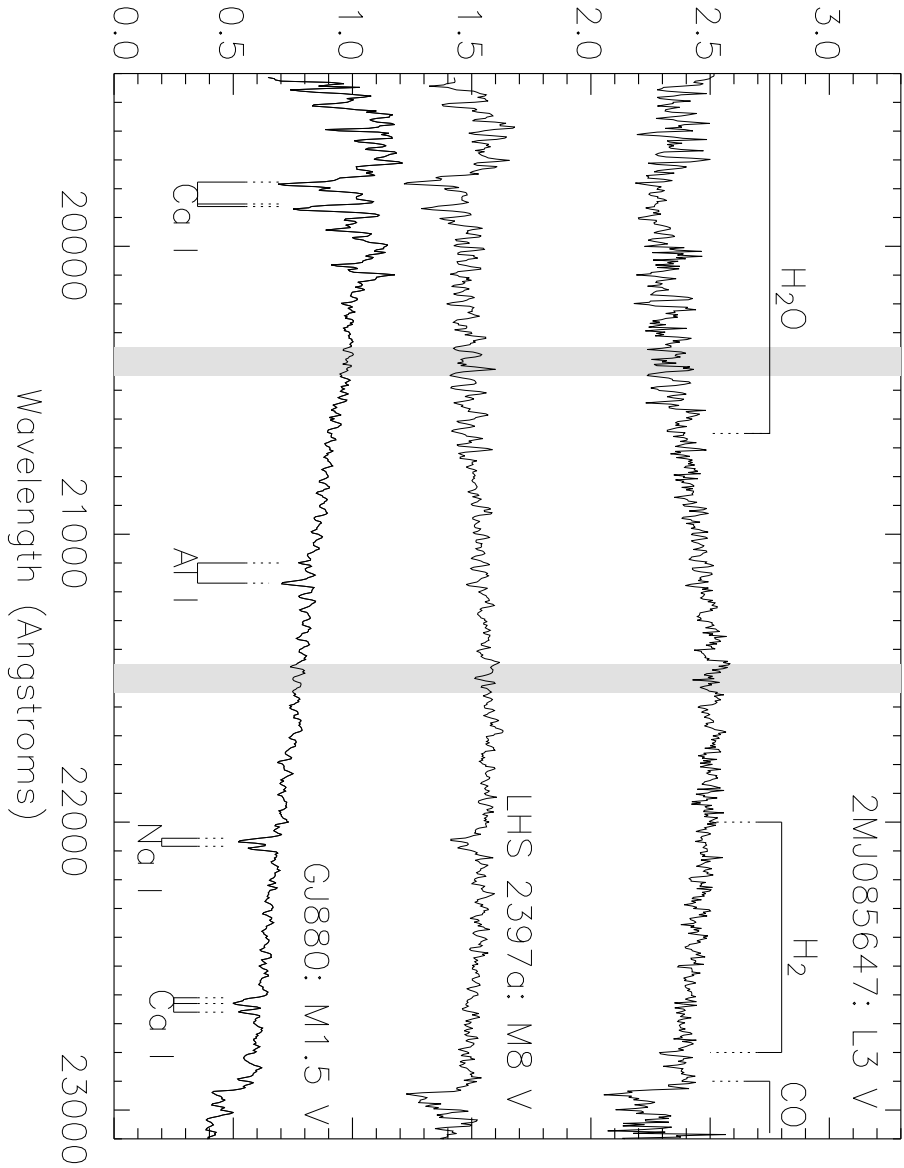}
\caption{$K$-band spectra of standard stars taken over a wide range in temperature. 
Atomic lines (CaI, AlI and NaI) and molecular features (H$_2$O, H$_2$ and CO) are labeled.
Flux bands centered on H$_2$O
and continuum features used in classification are shown as shaded regions.}
\label{fig:ktemp}
\end{figure}

Figure~\ref{fig:ktemp} shows a sequence of $K$-band M \& L dwarf standard spectra
taken with NIRSPEC.  We find the strong atomic line transitions 
(Ca I at 1.978 $\mu$m, 1.985 $\mu$m, 1.986 $\mu$m; 
Al I at 2.110 $\mu$m, 2.117 $\mu$m;  Na I at
2.206 $\mu$m, 2.208 $\mu$m; Ca I at 2.261 $\mu$m
, 2.263 $\mu$m \& 2.266 $\mu$m) all decrease in strength with temperature.  
The exceptions are the NaI lines which increase in strength until late M stars and then disappear quickly by early L
(see also McLean et al. 2003, 2000 and Reid et al. 2001).  The 1.98 $\mu$m Ca I triplet as well as the Na I and Al I lines decrease in 
strength with decreasing surface gravity but do not disappear
completely at early M spectral types, even in giant stars (Kleinmann \& Hall 1986).
The dominant molecular features are H$_2$O (1.90 $\mu$m), H$_2$ (2.20 $\mu$m) and CO (2.30 $\mu$m).

It is common practice when classifying low mass stars from $K$-band spectra to use the strength of the 2.30 $\mu$m CO absorption feature.
Because the K' filter used in our NIRSPEC setup begins to cut out at $\lambda \sim 2.295 \; \mu$m, we find the depth
of this feature in our data to be unreliable and therefore do not use it in classification.  
Again, following the work of McLean et al (2000) and Reid et al (2001), we have constructed an index to measure the strength of the very broad 1.90 $\mu$m H$_2$O feature (which effects the spectra from the short wavelength
end of our spectral range through $\sim$2.1 $\mu$m).  We define our index, H$_2$O-2, as the 
ratio of the flux at 2.04 $\mu$m to that at 2.15 $\mu$m (shown as shaded regions
on Figure~\ref{fig:ktemp}).  
The indices have been shifted slightly from those of other authors due to
ONC nebular emission features.
A plot of spectral type vs. the H$_2$O-2 index is shown in the lower left panel of Figure~\ref{fig:jh2o}.
The strength of this index decreases with decreasing spectral type,
similar to index behavior in the $J$-band.
We find good correlation for objects later than $\sim$M2.  
We derived an empirical fit to the standard observations which we used to classify our program stars.  Typical errors are $\sim$1.5 spectral subtypes.
Filled
squares in both lower panels correspond to low surface gravity field giant
stars.  
Wilking et al. (2004) found their $K$-band water index 
(which incorporates both the 1.9 $\mu$m and 2.5 $\mu$m H$_2$O absorption features)
to be insensitive to surface gravity in the dwarf to subgiant range.  These results are 
consistent with those of Gorlova et al. (2003) who compared the water index for
spectral type M young cluster objects and a sample of M-type field stars.
However, Wilking et al (2004) do find significant scatter in the measure of 
this index for giant stars.  They attribute this result in part to 
the variable nature of late-type giants.  We draw a similar conclusion
for the water index of giant stars from our own, admitedly extremely limited sample
(three out of four of which are Mira Variables).  
Since we do not have a large grid of young standard stars from which to derive
our own assessment of surface gravity effects on $K$-band H$_2$O absorption,
we assume those of Wilking et al. (2004) and Gorlova et al. (2003) and 
apply the relationship found for dwarf stars to our sample of pre-main-sequence targets.

Finally, Tokunaga \& Kobayashi (1999) have shown H$_2$ absorption to be present 
at $\sim$2.20 $\mu$m in late M and early L dwarfs. 
They define an index,
\begin{displaymath}
K2 = \frac{<F_{2.20-2.28}>-<F_{2.10-2.18}>}{0.5\left(<F_{2.20-2.28}>+<F_{2.10-2.18}>\right)}.
\end{displaymath}
where $<F_{\lambda_1-\lambda_2}>$ is the average flux between $\lambda_1$ and
$\lambda_2$.  We use a similar index, H$_2$, using the integrated flux between
$\lambda_1$ and
$\lambda_2$.
A plot of spectral type vs. H$_2$ is shown in the lower right panel of 
Figure~\ref{fig:jh2o}.
Because the dynamic range of this index is small and the scatter significant,
we do not derive an empirical fit.
However, the presence of H$_2$ absorption is readily detected in spectra
of late M and early L objects through visual inspection and was used 
in classification.  For reference, we have included measurements of
H$_2$ in Table 1.

\subsection{CRSP Spectra}
All $J$ \& $K$-band ONC spectra taken with CRSP were classified using the spectral 
features discussed
above.  Differences in equipment and data reduction techniques caused 
an offset in the continuum shape as compared to the NIRSPEC data and therefore we could 
not use the same band index relations.
Because we had CRSP spectra for only 16 
ONC objects, we relied on visual inspection for classification.  Extinction 
estimates were made 
in the same manner as for the NIRSPEC data (see Section 3.4) and object spectra
were compared to artificially reddened spectra of standard stars taken with CRSP
along with the target objects. 
Spectral types for objects inside and outside the inner 5.'1$\times$5.'1 of the ONC
are given in Table 1a and 1b, respectively.  
All objects have optical classifications either in the literature or from
new LRIS spectra
and these are also listed.  

\subsection{Extinction}

\begin{figure}
\epsscale{0.8}
\plotone{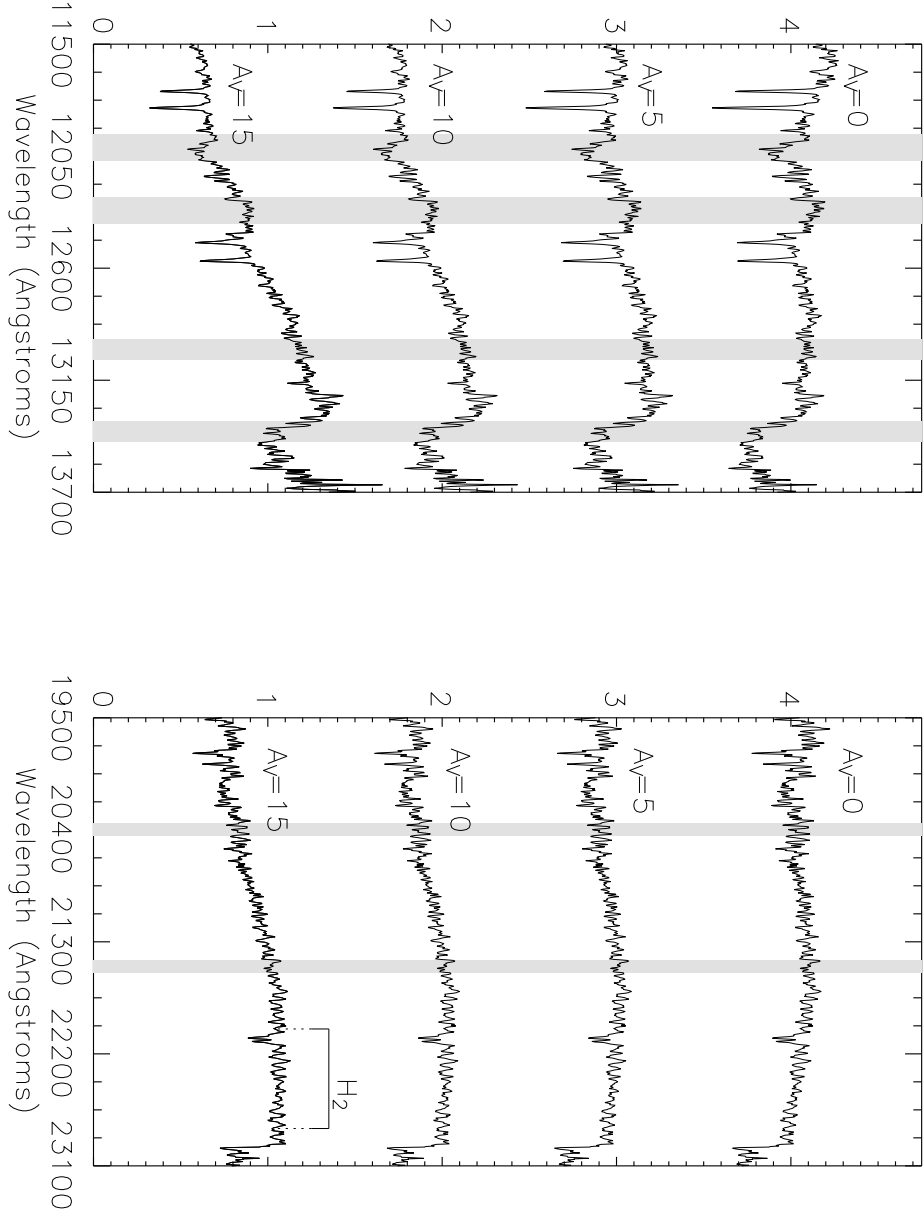}
\caption{M9 main-sequence standard star (LHS 2065)
which was observed during both NIRSPEC observing runs ($K$ \& $J$-Band).
In each panel, the top spectrum is the original data.  
Subsequent spectra have been artificially reddened by 5, 10 
and 15 magnitudes of visual extinction.
The flux bands corresponding to classification indices described in the text 
are shown as shaded regions and the $K$-band H$_2$ absorption region is marked.  We expect a systematic shift in all of the indices with increased reddening.}
\label{fig:redspec}
\end{figure}

As mentioned in the previous sections, the indices we use to derive spectral 
types are sensitive to effects from extinction which differentially reddens the 
spectra, thus, changing the band-to-continuum ratios.
Figure~\ref{fig:redspec} shows an M9 main-sequence standard star (LHS 2065)
which was observed during both NIRSPEC observing runs ($J$- \& $K$-band).
In each panel, the top spectrum is the original data.  
The subsequent spectra have been artificially reddened by 5, 10 
and 15 magnitudes of extinction using the IRAF $deredden$ package (which incorporates
the empirical selective extinction
function of Cardelli, Clayton \& Mathis, 1989).  The flux bands for the classification indices described in Section 3.1
are shown as shaded regions and the $K$-band H$_2$ absorption region is marked.  
We expect a systematic shift in
all of the indices with increased reddening.

\begin{figure}
\epsscale{1.0}
\plotone{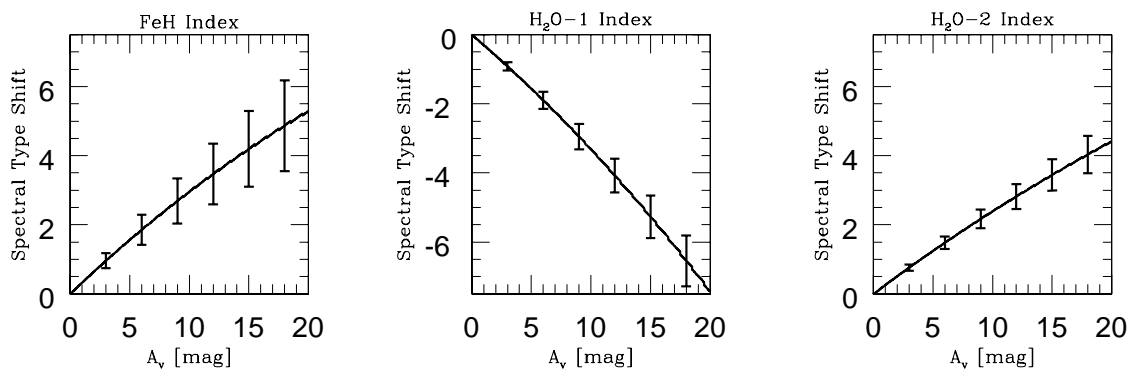}
\caption{Spectral type shift produced from index measurements as a function of A$_V$.  Extinction causes us to systematically classify a star later than
it is using the FeH and H$_2$O-2 indices, and earlier than it is using the H$_2$O-1 index.
Error bars correspond to 
errors in the fit, which increase with increasing A$_V$.  We find 
an average value of
A$_V$ $\sim$6 for our objects, 
which would result in a spectral type shift of $\sim$2 sub-types
for all indices if we did not attempt to take extinction into account
during the classification process.}
\label{fig:rband}
\end{figure}

To determine the magnitude of this effect, we artificially reddened all standard
objects by 1-20 magnitudes and re-measured each band index.  
We then derived empirical fits of the shift of each classification index as a function of increasing extinction.
Figure~\ref{fig:rband} illustrates the resulting spectral type shift 
plotted as a function of A$_V$ for all three indices.
Extinction causes us to systematically classify a star later than
it is using the FeH and H$_2$O-2 indices, and earlier than it is using the H$_2$O-1 index.
Error bars correspond to 
errors in the fit, which increase with increasing A$_V$.  
As can be seen, extinction must be taken into consideration when applying
the indices.
We find 
an average value of
A$_V$ $\sim$6 for our objects, 
which would result in a spectral type shift of $\sim\pm$2 sub-types
for all indices if not accounted for. 

We use color information to derive an initial extinction estimate for 
each of our spectral sources.  It has been shown (see Hillenbrand et al. 1998, Lada et al. 2000, and Muench et al. 2001) 
that $\sim$50\% of sources in the ONC show signs of ($J-H$),($H-K$) infrared excess
consistent with emission from a circumstellar disk.  Assuming the ($H-K$) excess
to arise from extinction alone without taking into account infrared excess
results in overestimates of A$_V$.  
Meyer, Calvet, \& Hillenbrand (1997) showed that the intrinsic colors of late-K and early-M 
stars 
with disks are confined to a classical T Tauri (CTTS) locus in the ($J-H$),($H-K$) color-color plane.  Following the work of M02,
we derive individual reddening $estimates$ for each star by dereddening its
($J-H$),($H-K$) colors 
back to this CTTS locus.  Stars falling below this locus (primarily
due to photometric scatter) are assumed to have an A$_V$=0.
For 12 of the objects which have no $J$ magnitudes available, 
A$_V$'s were estimated by dereddening $K,(H-K)$ magnitudes and colors 
back to a theoretical 1 Myr isochrone.   

Because the HC00 data set does not include $J$--band photometry, we used the color data of M02 
for de-reddening purposes.  In comparing the two photometric data sets, M02 \& HC00, we derived 
1 $\sigma$ standard deviations of $\sim$0.35 for the $H$ \& $K$ magnitudes.  
The 
scatter appears to be random for the $K$ magnitudes, but M02's $H$
photometry is systematically brighter than HC00's.  
Scatter between the two data sets 
was also noted by M02 who attributed the difference, 
in 
part, to the intrinsic infrared variability of low mass PMS stars.  Variables 
in the Orion A molecular cloud have been
found to have an average amplitude change of $\sim$0.2 mag at near-IR wavelengths 
on weekly to monthly scales (Carpenter, Hillenbrand \& Strutskie 2001). M02 also
noted that because of the
strong nebular background, differences in aperture sizes will contribute to
photometric noise.


For each program object we apply 
the empirical fits shown in Figure~\ref{fig:rband} to 
de-redden the measured spectral typing indices, 
and then use the new value to derive
an estimate of spectral type.  We compare each spectrum to standard stars,
artificially reddened to the same extinction value.  All spectral types
were verified visually to avoid contamination of band index measurements 
arising from noise in the spectra.

\begin{figure}
\epsscale{0.8}
\plotone{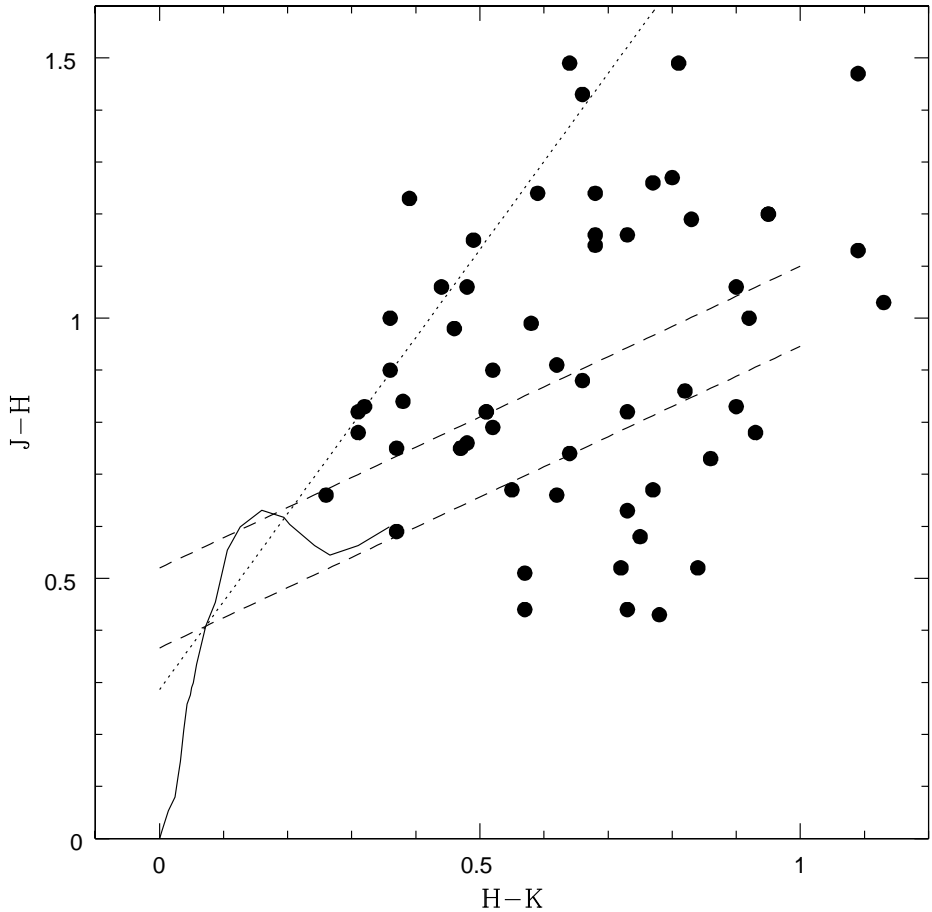}
\caption{Color-color diagram of ONC stars for which we have new near-IR 
spectral types later than K7. Data are
taken from M02.  The solid line represents the intrinsic colors of main-sequence dwarf
stars as given by Bessell \& Brett (1988), transformed to the CIT photometric system.
The slope of the interstellar reddening vector (dotted line) is that of Cohen et al. (1981).  Dashed lines indicate the upper and lower boundaries of the CTTS locus (Meyer, Calvet \& Hillenbrand, 1997) shifted to apply to
K7-L3 dwarfs.  
Note that we have not accounted for the intrinsic 
width of the locus as it is defined, or any 
possible change in slope of the locus with later spectral types.
If attributed to reddening, the width of the region corresponds to $\Delta$A$_V$ $\sim$2, or
less than one spectral subtype (see Figure~\ref{fig:rband}).}  
\label{fig:ctt}
\end{figure}

A possible bias in the above analysis is that the CTTS locus of Meyer, Calvet
\&  Hillenbrand (1997) applies
to the Taurus cluster, where most of the stars have spectral types 
K7-M0.  When applying the locus to stars outside this spectral range, 
we must take into account its anticipated width.  
Figure~\ref{fig:ctt} is a ($J-H$),($H-K$) color-color diagram for all of our
ONC stars with spectral types (excepting 12 stars for which there was no
$J$-band data available).  The top dashed line is the CTTS locus 
for Taurus stars, which defines an upper bound to the locus
when we consider all stars with spectral types K7-L3.  The bottom dashed line 
defines a lower bound to this locus.
The width of the region corresponds to a $\Delta$A$_V$ of $\sim$2, or
$\sim$0.5 spectral subtype uncertainty (see Figure~\ref{fig:rband}).  
Since intrinsic
spectral type errors derived from the indices are $\approx$1.5--2 subtypes, ignoring this effect does not result in significant biases.
In referring to the width of the locus we have only considered how the 
CTTS line as defined by Meyer, Calvet \& Hillenbrand (1997) would shift
if it applied to stars with K7-L3 spectral types; 
we have not accounted for the intrinsic 
width of the locus as it is defined, nor for any 
possible shift in slope of the locus with spectral type.
We emphasize that we are using this method to derive an A$_V$ $estimate$ 
only, to aid us in spectral classification.  
For placement
of a star on the HR diagram 
more precise
extinction values are derived 
by combining a star's photometric and spectroscopic
data (see Section 4.1).

\subsection{Infrared Excess}

As mentioned previously, we expect $\sim$50\% of the ONC sources to 
have a wavelength-dependent infrared 
excess due to thermal emission from warm dust grains in a circumstellar disk.
This excess dilutes (veils) the strength of molecular absorption features
used in spectral classification.
From experiments with artificially veiled standard stars, we find that
veiling can make a later-type spectrum look like an earlier-type photosphere. 
However, since the photospheric flux of late-type stars peaks at 1-1.5 $\mu$m, we 
do not expect extremely large excesses at near-IR wavelengths.  
The veiling index is defined as $r_\lambda=F_{_\lambda{excess}}/F_{_\lambda{photosphere}}$.
Meyer, Calvet \& Hillenbrand (1997) found that K7-M0 classical T Tauri stars have a median 
veiling value of
$r_K \approx$0.6.  We expect this value to be lower for cooler objects
whose photospheric emission peaks at longer near-infrared wavelengths.

\begin{figure}
\epsscale{0.8}
\plotone{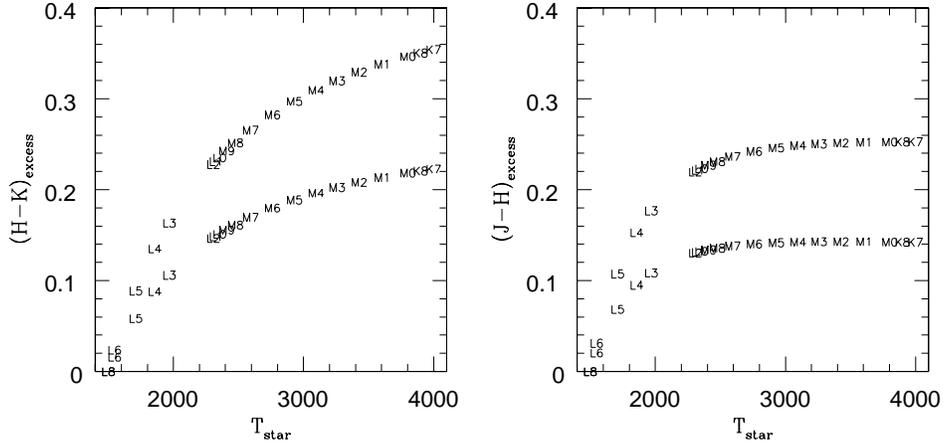}
\caption{Expected color excess for low mass stars veiled
by a T=1500 K blackbody.  
Data points are labeled by spectral type and the two rows correspond to 
$r_K$ = 0.5 (bottom)  and 1 (top). 
For $r_K <$ 0.5, we expect 
color excesses $\Delta(H-K)$ and $\Delta(J-H)$ $\lesssim$ 0.2 mag.  
The effect of veiling on classification is $\approx $1.5--2 spectral subclasses earlier 
for the
H$_2$O-2 and FeH indices and later for the H$_2$O-1 index.
}
\label{fig:veil}
\end{figure}

Fig~\ref{fig:veil} shows the expected color excess for stars of varying
spectral type veiled
by a T=1500 K blackbody.  
Data points are labeled by spectral type and the two rows correspond to 
$r_K$ = 0.5 (bottom) \& 1 (top).  
As can be seen,
for $r_K <$ 0.5, we expect 
color excesses $\Delta(H-K)$ \& $\Delta(J-H)$ $\lesssim$ 0.2 mag.  
The effect on classification is $\approx $1.5--2 spectral subclasses earlier/later 
for the
H$_2$O-2 \& FeH/H$_2$O-1 indices.
This result is somewhat over-estimated for later spectral types 
(cooler than $\sim$M6) with $K$-band spectra for which it is more readily apparent via visual inspection when
veiling is affecting spectral classification.  Because we have no way to 
independently measure veiling in our spectra
(modeling of high dispersion data is required), we simply note the possible bias.

\subsection{Summary of Spectral Classification}

In all cases, classification was done first  
using flux ratios of broad molecular absorption 
lines to continuum, then comparing visually to standards.  Extinction
was estimated assuming CTTS colors and taken into account during the classification process. The influence of veiling was investigated but not explicitly
accounted for during spectral classification.
Surface gravity was assessed visually from the presence/strength of atomic 
absorption lines.  For $J$-band spectra an object was given a gravity classification of ``low" if it had no detectable
atomic absorption lines and ``high" if it had strong lines, similar to those seen in spectra of dwarf standards
of the same spectral type.
A classification of ``int" indicates absorption lines were present but not as strong as those in dwarf stars at the 
same temperatures (see Table 1a).  For $K$-band spectra gravity determinations
were given based on the relative strength of atomic absorption features.

From the NIRSPEC data, we have classified 71 objects in the inner ONC
found to be of spectral type K7 or later,
$\approx$50\% of which are M6
 or later.    
At an age of 1-2 Myr, all objects with spectral types later than M6 are substellar
(M $<$ 0.08 M$_\odot$)
based on comparison with theoretical models (e.g., DM97).
In addition, we have classified 16 spectra of objects in the 
inner and outer parts of the nebula taken with CRSP, 9 of which are M6 or later.
Finally, we also report in Tables 1a \& 1b new optical spectral types obtained with
LRIS, as well as those reported previously in the literature.
For the large majority of sources there is excellent agreement (to $<$ 2 
spectral subclasses) 
between optical and infrared spectral types.

\section{HR Diagram}

In this section we combine each object's spectral type and near-IR photometry 
to derive values for its luminosity and effective temperature and place it
on the theoretical HR diagram.
The goal is to transform temperatures and luminosities into ages
and masses using theoretical pre-main sequence evolutionary tracks.
From this information we can explore the ONC's IMF down to $<$ 0.02 M$_\odot$
using spectroscopically-confirmed cluster members.

\subsection{Effective Temperatures, Intrinsic Colors \& Bolometric Luminosities}

Spectral types of K7-L3 were converted to temperatures based on the scales of
Wilking, Green \& Meyer (1999), Reid et al. (1999) and Bessell (1991). 
An empirical fit for BC$_K$ was determined from the observational data of
Leggett et al. (1996, 2002) (spectral types M1-M6.5, M6-L3).
We determined intrinsic colors by computing empirical fits to a combination
of Bessell \& Brett (1988) theoretical models (for spectral types K7-M6)
and Dahn et al. (2002) observed data (spectral types M3.5-L3).
$(H-K)$ color-excesses were converted to A$_K$ using the reddening law of
Cohen et al. (1981).
Because $J$ magnitudes are not available for all program stars, for consistency 
we used $(H-K)$ colors and $K$ magnitudes from HC00 to determine extinction
values and estimate bolometric corrections. 
Using ($H-K$) colors rather
than ($J-H$) colors to predict extinction results in higher derived
A$_V$ estimates by an average of $\sim$3 mag for our data ($\sim$0.2 mag in A$_K$).  
In addition, if the object has substantial infrared excess, its 
$K$-band magnitude will be brighter than the photosphere.  
However, the combination of these effects 
does not produce a significant shift in placement of a star on the HR diagram.
We find an average difference of
$\Delta \log{L} \approx$0.05 between luminosities derived
using $J,(J-H)$ vs. $K,(H-K)$ for objects with both
$J$- and $K$-band data available.  A similar trend of comparable magnitude 
results if we compare A$_V$ values
derived from ($H-K$) colors alone rather than using both ($J-H$) and ($H-K$)
to deredden stars back to the CTTS locus (see Section 3.4).  
The uncertainty in derived A$_V$ for an M6 star with large color and 
spectral type uncertainties ($\sigma_{(H-K)} \sim0.05$ mag and $\sigma_{Spec Type} \sim 2$ subclasses)
is $\approx$1.5 magnitudes.  
All fits are given in Table 2 and  
derived quantities for stars included in the
HR diagram are given in Table 3.

For all derived quantities discussed above, we used dwarf scales despite the lower surface
gravity of young pre-main-sequence stars.  It has been known since the earliest work on T Tauri stars
(Joy 1949) that young, pre-main-sequence objects are observationally much closer to dwarfs than to
giants or even sub-giants.  
Since 
to date there is no accurate temperature or bolometric correction scale
for pre-main-sequence stars, we use the higher gravity dwarf scales
in our analysis. 

\subsection{HR Diagram for objects with NIRSPEC data}
\begin{figure}
\epsscale{0.8}
\plotone{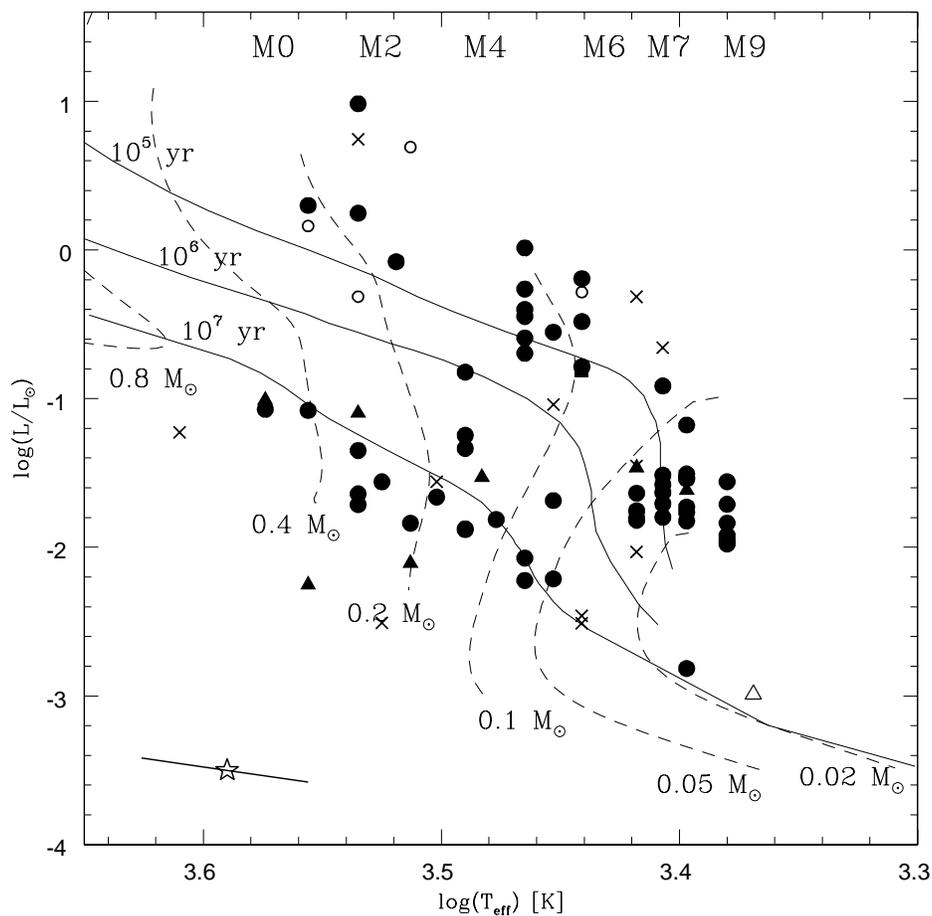}
\caption{HR diagram of objects
within the inner 5.'1 $\times$ 5.'1 of the ONC (survey area of HC00)
for which we have new
spectral types.  Filled 
symbols correspond to stars classified using
NIRSPEC and CRSP infrared observations which have good
spectral types.
Objects which were classified as ``high"
gravity (i.e., had strong absorption lines, similar in strength to 
dwarf stars of the same temperatures) are marked as open symbols and
objects with less certain spectral types (those designated with a ``:''
in Table 1a) are marked as X's.
Triangles represent objects which have new optical spectral types
taken with LRIS.
A typical error bar for an M6$\pm$1.5 star is shown in the lower left corner.
The pre-main-sequence model tracks and isochrones of DM97 
are also shown.}  
%
\label{fig:hrd}
\end{figure}

In Figure~\ref{fig:hrd} we present an HR diagram for those objects
within the inner 5.'1 $\times$ 5.'1 of the ONC (survey area of HC00)
for which we have new
spectral types.  Objects which were classified using spectra from different
instruments are plotted as different symbols.  Surface gravity assessment is
indicated and   
a typical error bar for an M6$\pm$1.5 star is shown in the lower left corner.
The pre-main-sequence model tracks and isochrones of DM97 
are also shown.  
Figure~\ref{fig:hrd} illustrates that we are able to probe lower masses
than previous spectroscopic studies have, down to 
0.02 M$_\odot$.

From examination of Figure~\ref{fig:hrd}, it appears that we are exploring
two separate
populations: the inner ONC at $\lesssim$ 1 Myr, and an older population
at $\sim$10 Myr.  
None of the members of this apparently older population have gravity features
which indicate they might be foreground M stars.  Spectra of these objects indicate instead that they have
low surface gravity consistent with young objects,
and are therefore also candidate 
cluster members.
For the remainder of Section 4.2 we will discuss the possible explanations
for this surprising population.  
First, we detail possible systematics in the data 
reduction/analysis which could, in principle, cause one to 
``create'' older stars in the HR diagram.  Next we discuss their possible origin
if they are a real feature and finally we determine why they may not have
been detected in previous studies.

There are two possible reasons we might erroneously detect 
a bifurcation of the HR diagram:
either the spectral types are in error and the objects are actually cooler
than they are shown to be here, or their luminosities have been
underestimated.
The photometric uncertainties for these objects are $<$ 0.2 mag for both 
$H,K$ in the HC00 data and $<$ 0.4 mag for $J,H,K$ in the M02 data.  The two data sets agree to within 0.8 mag for all objects and variability, if
present, is expected to be $\sim$0.2 mag (see Section 3.4).
An object with spectral type M2 would need to be 
$\sim$3 magnitudes brighter in $K$, or have its extinction underestimated by
$>$ 30 mag in order to move on the HR diagram from the
10 Myr isochrone to the 1 Myr isochrone.
Even taking into account the above uncertainties and possible variability,
errors of this order are not possible.  Moreover, if the photometry
were in error, we would expect the result to be continuous scatter in
the HR diagram, rather than two distinguishable branches.    

The other way to account for the apparently 
older population through error is if the temperature
estimates are too hot, i.e., the spectral types too early, in some cases by
as much as 7 spectral subclasses.  
All spectra have been checked visually by multiple authors (C.L.S \& L.A.H.)
to ensure accurate classification
and we believe our spectral types are robust to within the 
errors given.  
None of the possible biases resulting from assumptions made regarding veiling,
extinction or gravity can have a large enough effect to produce
a spectral type offset of this magnitude.  
One possible exception is that 
a mid-M (M5) K-band spectrum with high reddening can 
have the same appearance and band 
index measurements as a later-M (M7) spectra with intermediate reddening.
But this phenomenon could only affect a small subset of the 
lower branch population.  Therefore, while it may reduce slightly the number
of apparent older objects, it cannot correct for them entirely. 

The accuracy of our spectral types
is supported by the excellent agreement
of our near-IR
spectral classifications with existing optical spectral types (most agree to within 2 spectral subclasses; see Table 1a).
We find a significant fraction of the spectra taken
with LRIS (filled triangles on Figure~\ref{fig:hrd}) which
were classified in a completely independent manner from
the NIRSPEC and CRSP data also 
resulted in several apparently 10 Myr old stars.  
If the older population arises from systematics in the reduction or 
spectral classification of our infrared spectra, we would not
expect to find stars classified optically to fall on the same place in the 
HR diagram.

If we are to accept the bifurcation as a real feature in the HR diagram,
 several scenerios could, in principle, account for it.
The first is that we are seeing contamination from foreground field stars.
However, as mentioned, HC00 found the expected field star contribution to be 
only $\sim$5\% down to the completeness limit of $K$$\sim$17.5 (see Section 1).
We find the apparent low luminosity population to account for a much larger fraction ($\approx \frac{1}{3}$) of our sample.
The second possibility is that we are seeing contamination 
by a foreground population of 
M stars not from the field, but from the surrounding OB association. 
The ONC (also known as Orion subgroup Id) is neighbored by three 
somewhat older subgroups of stars
(Ia, Ib, \& Ic; see Brown, de Geus
\& de Zeeuw 1994 for a detailed description of each) which are located 
at distances ranging from $\sim$360--400 pc and having 
ages (derived from high mass
populations) from $\sim$2--11.5
Myr.  
The most likely subgroup from which we would see contamination in our data
is subgroup Ic, which is located along the same line of sight as the ONC.  
However, its members are thought to be only $\sim$2 Myr old
and therefore cannot account for the stars on the lower branch
in the HR diagram unless their age estimate is in error.
The only known part of the OB association which could be contributing $\sim$10 Myr old
stars to our study is subgroup Ia, which is estimated to be $\sim$11.4 Myr.  

While the known high mass population of the Ia subgroup does not extend spatially 
as far as the ONC,
if there has been dynamical relaxation, it is possible that its
lower mass
members may occupy a more widespread area than the OB stars. 
Brown, de Geus
\& de Zeeuw (1994) found the initial mass function for the subgroups to be
a single power law of the form $\xi$(log $M$) $\propto$ $M^{-1.7 \pm 0.2}$
based on the high mass population ($M$ $>$ 4 M$_\odot$).  Since
the lower mass members have not yet been identified or studied in detail, we extrapolate
the high-mass IMF to determine the total number of stars in subgroup Ia expected
down to 0.02 M$_\odot$ and find there should be $\approx$3500 members.
This number is an upper limit since we have not applied a Miller-Scalo turnover to
 the IMF but simply extended the
power law form.
The angular size of the Orion Ia subgroup as studied by Brown, de Geus
\& de Zeeuw (1994) corresponds to a linear size of $\sim$45 pc.
From this we can calculate a relaxation time for the cluster
\begin{displaymath}
t_{relax} = n_{relax} \times t_{cross} = \frac{8 \ln{N}}{N}\times \frac{R}{\sigma}.
\end{displaymath}
Assuming a gravitationally bound cluster and a velocity dispersion of $\sigma \sim$2 km s$^{-1}$ consistent with the 
ONC (Jones \& Walker 1988), we find a crossing time for the Orion Ia cluster of $\sim$11 Myr and a 
relaxation time of $\sim$600 Myr.  Clearly the cluster is
not yet dynamically relaxed and it is unlikely that its low mass population 
would have spread significantly past its higher mass members.  This calculation does not however, rule
out the possibility that the lower mass population
formed over a wider spatial area than the massive stars, given that primordial 
mass
segregation has been observed in other young clusters.  

From the same IMF extrapolation
we find the surface density
of 0.5-0.02 M$_\odot$ objects in subgroup Ia
to be $\sim$2.5 pc$^{-2}$.  
The areal extent corresponding to the angular size of our 
survey region (assuming a distance of 480 pc)
is $\sim$0.5 pc$^2$.  Assuming a constant
surface density across this area we would expect to 
find $<$ 2 members
of Orion Ia in the current work.  
Even if we ignore previous age estimates, assume all three of the subgroups
could be contributing to the observed lower branch of the HR diagram, and 
repeat the above calculation for subgroups Ib \& Ic, we would 
expect
to see $<$ 10 stars total.
Therefore, it is unlikely that the lower branch we see in the inner ONC
HR diagram is purely due to contamination
from surrounding populations.
We cannot rule
out that there may exist a foreground population of $\sim$10 Myr stars which is not associated with the ONC, but
which was missed by the Brown, de Geus
\& de Zeeuw (1994) work due to its dirth of OB stars.  However, the probability that this population would lie in exactly the
same line of sight as the ONC is small.

Another possible cause of the older branch of the HR diagram is scattered light
from circumstellar disks and envelopes.
Most of the objects in the lower branch of the HR diagram
have ($J-H$),($H-K$) near-infrared excesses consistent with their being
young objects surrounded by circumstellar material which   
could result in
their detection primarily in scattered light.  
If true,
 extinctions, and consequently luminosities for these objects would be
underestimated, making them appear older than the bulk of the population.
Similar arguments have been put forth by Luhman et al. (2003b) to explain
low-lying stars on the HR diagram of IC 348.  
One uncertainty in this argument is that if scattered light is responsible
for causing the apparently older population, it would have to be acting
on our observations in such a way so as to create a dichotomy of object luminosities
rather than a continuous distribution of stars
at unusually faint magnitudes. This scenerio could occur however, if the 
lower luminosities measured for this population are caused by a drop in observed 
light due to the objects
being occulted by a disk.
The relatively low luminosities of substellar objects would 
make such systems more difficult to detect and we 
may be observing the brighter end of a distribution of 
low mass scattered light sources.  In support of this argument, one of these sources is coincident 
with an optically identified {\it Hubble Space Telescope} ``proplyd'' (which is known to be a photoevaporating circumstellar disk), 
and two with disks
seen optically in silhouette only (Bally, O'Dell, \&
McCaughrean 2000).

\begin{figure}
\epsscale{0.8}
\plotone{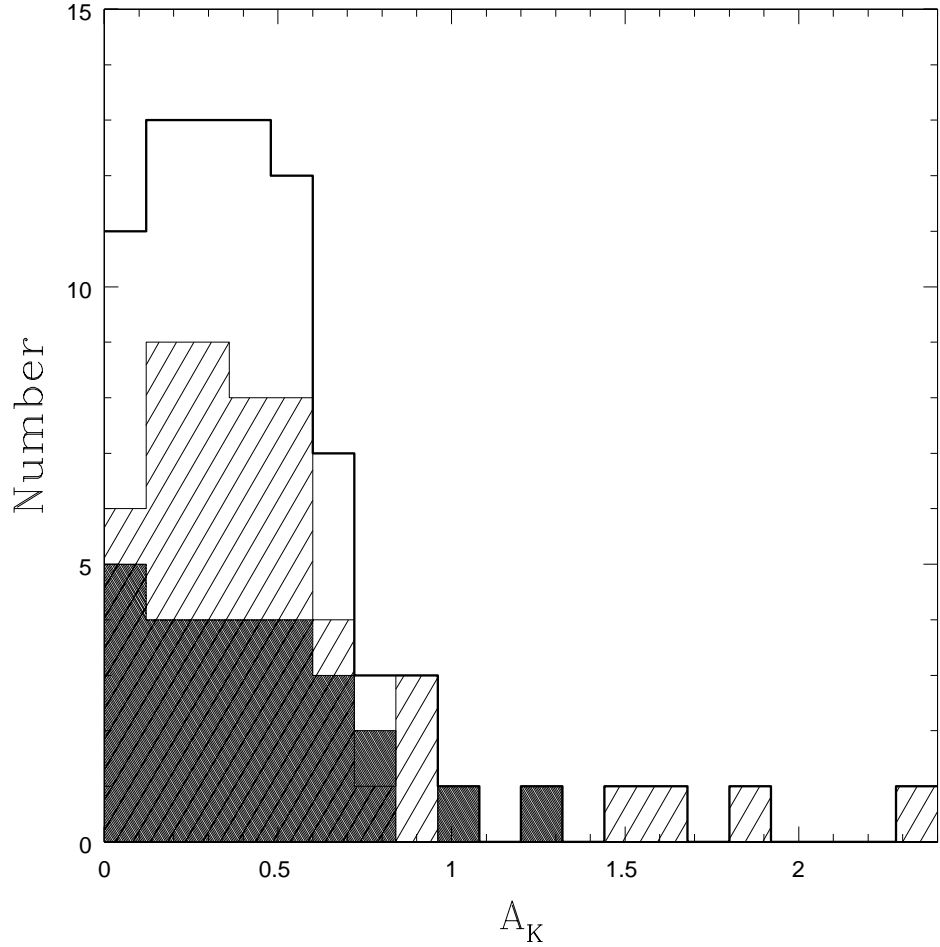}
\caption{Histogram of extinction for sources included in 
Figure~\ref{fig:hrd}.  The open histogram indicates all objects; 
the hatched and shaded histograms include the two populations with apparent ages (as interpretted from the HR diagram) 
of $\sim$1 and $\sim$10 Myr old, respectively.  
We find smilar extinction distributions 
for all three populations.}
\label{fig:av}
\end{figure}

If the lower luminosity population in the inner ONC 
is real, rather than due to systematics in the data reduction/analysis or contamination from the surrounding older association, 
the question arises as to why it was not seen in previous surveys.  The 
deepest optical photometric survey in this inner region is that by 
Prosser et al. (1994) which
is reported complete to V$\approx$20 mag and I$\approx$19 mag, though
no details are given on how these numbers were derived or how they might vary with
position in the nebula.  One can view the photometric data 
of H97, which is a combination of ground-based CCD photometry
out to 15' from the cluster center and the Prosser et al. (1994) HST CCD
photometry in the inner 3' or so, in terms of color-magnitude diagrams
binned by radial distance from the cluster center.  At all radial distances
there are some stars which sit low in the color-magnitude diagram.  However,
it is only in radial bins beyond $\sim$7' that a substantial population (though still
disproportionately small compared to our current findings)  
of apparently 10 Myr old low mass (0.1-0.3 M$_\odot$) stars begins to appear, 
as might be expected if they are contaminants from the Orion Ic association.  
The inner radial bins do not display this population, 
though it should have been detected at masses
larger than 0.15 $M_\odot$ $in\; the\; absence\; of\; extinction$.  
Notably the
apparently 10 Myr old population in Figure~\ref{fig:hrd} does not lack extinction.
Figure~\ref{fig:av} shows a histogram of extinction for sources included in 
Figure~\ref{fig:hrd}.  The open histogram indicates all objects; 
the hatched and shaded histograms include only the 1 and 10 Myr old
populations, respectively.  
We find all three populations to have similar, roughly constant distributions 
for A$_K \; \lesssim$ 0.6 (A$_V \; \lesssim$ 10)
giving further evidence that the lower luminosity branch 
of the HR diagram  is 
likely not a foreground population.  
Accounting for typical errors ($\sim$0.1 mag 
at $K$ and $\sim$1.5 mag at $V$; see Section 4.1) will not change our conclusions.   
For extinction
values larger than about 2 visual magnitudes the 0.15-0.5 M$_\odot$ stars at 10 Myr in
Figure~\ref{fig:hrd} would have been missed by the optical 
Prosser et al. (1994) photometric survey but uncovered in later infrared surveys.
Further, Walker et al. (2004) find from theoretical work a lower incidence of disk occultation for higher
mass CTTS ($\sim$20\% of systems with disks) in comparison to lower mass substellar objects 
(up to $\sim$55\% of systems with disks) due to smaller disk
scale heights and less disk flaring.  Therefore, if the low luminosity objects
are a population of scattered light sources we might expect to observe 
smaller numbers of them at higher masses.

\section{The Low-Mass IMF}

In this section we use theoretical mass tracks and isochrones to determine
a mass and age for each object in our HR diagram.  
In Section 5.1 we discuss our choice of models and in Section 5.2 
the different criteria we consider for including a star in the IMF.
Finally, in Section 5.3 
we present the derived mass function and point out important features.

\subsection{Models and Mass Estimates}

Currently, there are relatively few sets of pre-main-sequence (PMS)
evolutionary models which extend into the substellar regime. 
The Lyon group (see Baraffe et al. 1998 \& Chabrier et al. 2000) 
calculations cover 0.001--1.2 M$_\odot$, but do not extend to the 
large radii of stars younger than 1 Myr.  Thus, they must be applied 
to young low-mass star forming regions with caution.
The other widely utilized set of low-mass PMS models are those by 
DM97.
These models cover 0.017--3.0 M$_\odot$ over an age range of 
7$\times$10$^4$ yr to 100 Myr, and are therefore representative
of even extremely young regions.  A detailed analysis and comparison
of the models is given by Hillenbrand \& White (2004).
For the purpose of the current work we use DM97 tracks and isochrones
to determine masses and ages for stellar and substellar objects
in our HR diagram.

\subsection{Completeness}

\begin{figure}
\epsscale{0.8}
\plotone{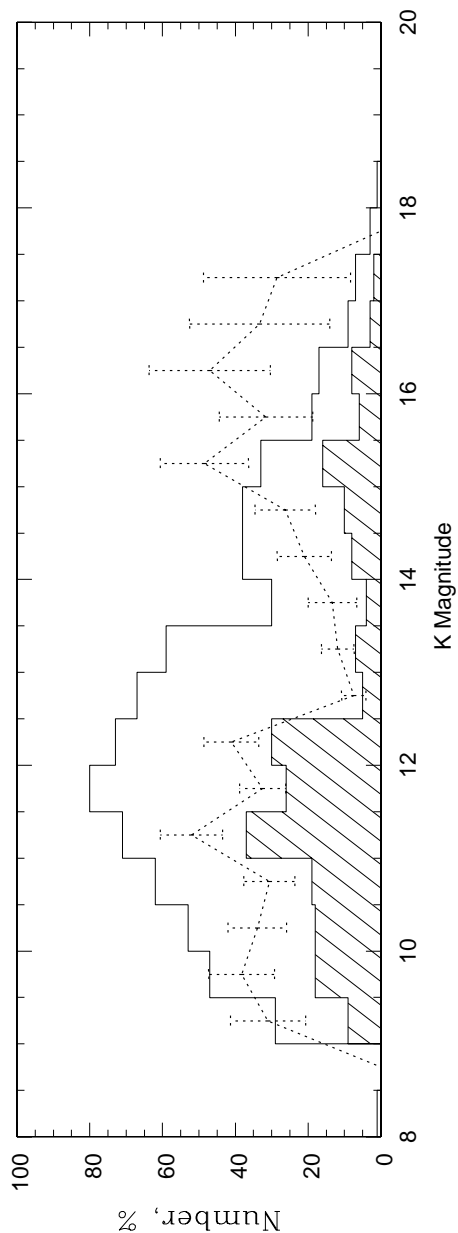}
\caption{Histogram of HR diagram completeness as a function of magnitude.
Open and hatched histograms indicate the distribution of all 
HC00 photometry and the subset 
for which we have spectral types, respectively.  The dotted line
in each panel indicates fractional completeness of the spectroscopic sample with $\sqrt(N)$ errorbars.
Although the spectroscopic sample is $\approx$40\% complete at bright and faint magnitudes, correction
factors to the mass function are needed at intermediate magnitudes.}
\label{fig:maghist}
\end{figure}

Because the ONC is highly crowded and extremely nebulous, we must 
be certain that the spectroscopic sample is representative of the 
population as a whole.  
Figure~\ref{fig:maghist} shows a 
histogram of completeness as a function of magnitude.
Open and hatched histograms indicate the distribution of HC00 photometry and of
sources for which we have spectral types, respectively.  
The hatched histogram includes not only sources for which we
have new spectral types, but also optically-classified sources from the literature
(H97 and references therein)
which fall within our survey region and were determined to be of spectral type
M0 or later.  
In combining the two data sets our goal is to assemble a statistically
large sample of cool, spectroscopically confirmed ONC members
down to extremely low masses (M $\sim$0.02 M$_\odot$) from which to create
our IMF.   
 
The dotted line
indicates fractional completeness with $\sqrt(N)$ errorbars.
We estimate that we are $\sim$40\% complete across our magnitude range with the 
exception of the 12.5 $<$ $K$ $<$ 15 bins.  
Undersampling at these intermediate magnitudes 
is caused by our combing the sample of brighter, optically-classified, 
mainly stellar objects
with our sample of infrared-classified objects where the aim was to 
observe objects fainter than the substellar limit.  Therefore, magnitude bins
corresponding to masses near the hydrogen burning limit are underrepresented.
We can correct for this by determining the masses which correspond to objects in 
the depleted magnitude bins.  We then add stars to those mass bins according to 
the fractional completeness of the magnitude bins they came from, relative to 
the overall 40\% completeness.  We discuss both the corrected and non-corrected IMFs in the next section.

\subsection{The Stellar and Substellar IMF}

\begin{figure}
\epsscale{0.8}
\plotone{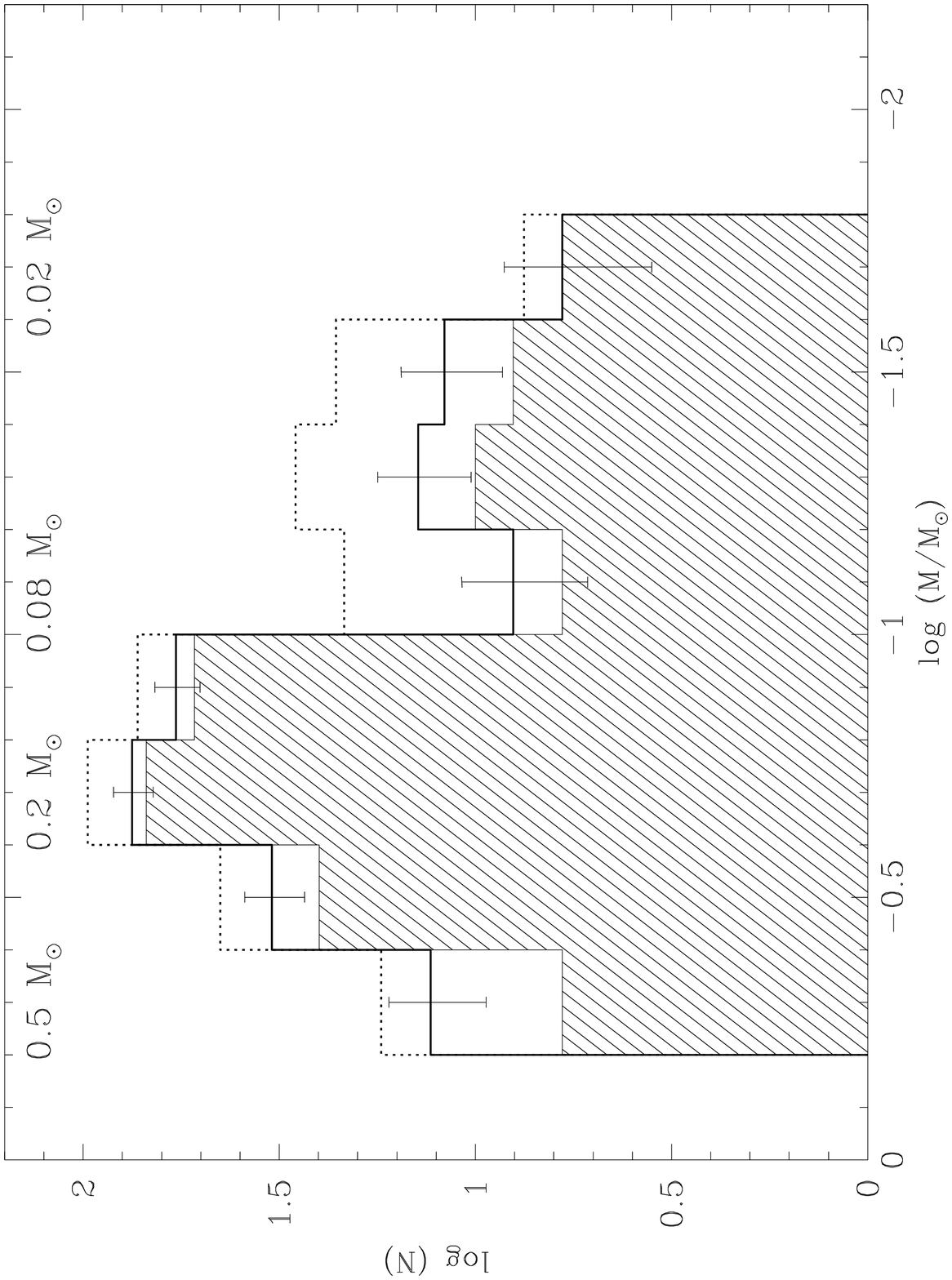}
\caption{Mass function for all stars within the inner 5.'1 $\times$ 5.'1
of the ONC that have spectral types from the present work or the literature
later than M0.  Thick-lined open histogram shown with $\sqrt{N}$ errorbars 
indicates all stars; the dotted
open histogram represents the same sample corrected for 
incomplete magnitude bins (Figure~\ref{fig:maghist}). 
The hatched histogram 
indicates only stars younger than $\sim$5 Myr according to the HR diagram (Figure~\ref{fig:hrd}).}
\label{fig:mf2}
\end{figure}

Figure~\ref{fig:mf2} shows the mass function for stars within the inner 5.'1 $\times$ 5.'1
of the ONC that have spectral types derived from either 
infrared or optical spectral data 
later than M0.
  The thick-lined open histogram, shown with $\sqrt(N)$ errorbars, 
includes all stars meeting this
criteria which were not determined to have high surface gravity (see Table 1a).
In total, the IMF sample includes $\sim$200 stars and contains objects 
with masses as high as 
 $\sim$0.6 M$_\odot$ and as low as $<$0.02 M$_\odot$.
Mass completeness (to the 40\% level)
extends from $\sim$0.4 M$_\odot$ 
(corresponding to an M0 star at 1 Myr; see Figure~\ref{fig:hrd}) to 
the completeness limit of HC00, $\sim$0.03 M$_\odot$ for A$_V$ $<$ 10. 
The dotted open histogram indicates an IMF for the same sample, where we
have corrected for the incomplete magnitude bins.
The most substantial correction occurs right at the substellar limit as
indicated in Section 5.2.
The hatched histogram represents only those objects 
determined to be younger 
than 5 Myr from HR diagram analysis.  Thus we remove any possible bias occurring
by inclusion of the apparently older $\sim$10 Myr 
population which may or may not be a real part of the cluster.
As can be seen, the apparently older stars constitute a relatively uniformly spaced  
population across the mass range we are exploring (see also Figure~\ref{fig:hrd}), and including them in the IMF 
does not affect our interpretation.

The stellar-substellar IMF for the inner ONC in Figure~\ref{fig:mf2} peaks
at $\sim$0.2 M$_\odot$ and falls across the hydrogen-burning limit
into the brown dwarf regime.  
Using the observed IMF we find $\sim$17\% (39) of the objects in our sample to be substellar
($M$ $<$ 0.08 M$_\odot$). 
Below $\sim$0.08 M$_\odot$ the mass
distribution levels
off through our completeness limit of $\sim$0.02 M$_\odot$.
There may be a secondary peak at $\sim$0.05 M$_\odot$. 
Decreasing the bin size by 50\% yields the same results, with the 
peak at 0.05 M$_\odot$ becoming even more pronounced.  Increasing the binsize 
by 50\% produces a steady decline into
the substellar regime.  The total mass inferred between 0.6--0.02 M$_\odot$ from our data is $\sim$41 M$_\odot$ which, corrected for our 40\% completeness
factor becomes $\approx100$ M$_\odot$.  This value is a lower limit because
we have already shown that we are not 40\% complete at all magnitudes.  
We find an average mass of $\sim$0.18 M$_\odot$ corresponding to the peak 
in our data.

\section{Discussion}

\subsection{Comparison to previous ONC IMF determinations}

The stellar/substellar IMF has been discussed in previous work on the ONC.
A determination based on near-infrared photometric data was made by HC00 using
$H$ and $K$ magnitudes and colors combined with star count data to 
constrain the IMF down to $\sim$0.03 M$_\odot$.  They find a mass function
for the inner regions of the ONC 
which rises to a peak around 0.15 M$_\odot$ and then 
declines across the hydrogen burning limit with slope N($\log M) \propto$ $M^{0.57}$.  M02 transform the inner 
ONC's $K$-band luminosity function
into an IMF and find  
a mass function which rises with decreasing mass to form a broad 
peak between 0.3 M$_\odot$ and the hydrogen-burning limit before turning over and falling off into the substellar regime.  This decline is broken between 0.02
and 0.03 M$_\odot$ where the IMF may contain a secondary peak near the 
deuterium burning limit of $\sim$13 M$_{Jup}$.
Luhman et. al (2000) combined near-infrared NICMOS photometry of the inner
140" $\times$ 140" of the ONC with
limited ground-based spectroscopy of the brightest objects ($K <$ 12) 
to determine a mass function which follows a power-law slope
similar, but slightly steeper than the Salpeter value, until it turns over
at  $\sim$0.2 M$_\odot$ and declines steadily through the brown dwarf regime.
H97 presents the most extensive spectroscopic survey 
of the ONC, combining optical spectral data 
with $V$ and $I$-band photometry over a large area ($\sim$ 2.5 pc$^2$) extending
into the outer regions of the cluster.  The IMF determination covers
a large spectral range and appears to be rising from the high (50 M$_\odot$)
to low (0.1 M$_\odot$) mass limits of that survey.

The peak in our IMF, $\sim$0.2 M$_\odot$, matches remarkably well to those 
found from both the deep
near-infrared imaging IMF studies (HC00 and M02) 
which cover similar survey areas, 
and the Luhman et al. (2000) study which covered only the very
inner region of the cluster.
Our data also show a leveling off 
in the mass distribution through the substellar
regime similar to that found by HC00. 
A significant secondary 
peak within the substellar regime 
has been claimed by M02.  While we see some evidence for such a peak
in our data, this result is not robust to within the errors.  Furthermore,
if real, we find the secondary peak to occur at a slightly higher
mass than M02 ($\sim$0.05 M$_\odot$ vs. $\sim$0.025 M$_\odot$). 
The primary difference between the observed IMF derived in the current work and those
presented in previous studies of the substellar ONC population
is the steepness of the primary peak 
and the sharpness of the fall-off at the hydrogen-burning limit 
(see for comparison Figure 16 of M02).
Most IMF determinations for this region exhibit
a gradual turnover in the mass function around $\sim$0.2 M$_\odot$ 
until 
$\approx \frac{2}{3}$ the level of 
the primary peak is reached at which point the IMF levels off or forms a secondary peak.
However, we find a sharp fall-off beyond $\sim$0.1 M$_\odot$ 
to $\approx \frac{1}{2}$ 
the primary peak value.  

\subsection{Photometric vs. Spectroscopic Mass Functions}

As mentioned in Section 1, spectroscopy is needed to study 
a cluster's IMF in more than a statistical sense. 
The fact that 
many of the fainter objects in our survey expected
to be substellar based on their location in the $K$, ($H-K$) diagram
are in fact hotter, possibly older objects gives strong evidence in support of this statement.  
Independent knowledge of an individual star's age, extinction and infrared excess arising from the possible presence of a circumstellar disk is needed
before definitive conclusions can be drawn about that object's mass.
Despite the increasing numbers of brown dwarfs studied, both in the field and
in clusters, details of their formation process remain sufficiently poorly
understood that knowledge of near-infrared magnitudes and colors alone
is not enough to accurately predict these characteristics for young 
low mass objects.  Near-infrared magnitudes alone also cannot 
distinguish between cluster members and nonmembers, and 
field star contamination must be modeled rather than accounted for directly.

Previous studies of the ONC had spectroscopy available in general only for the stellar
population, and relied on photometry alone to determine
the mass function at substellar masses.
This may have caused  over-estimates in the number of brown dwarfs 
for reasons discussed above.  
However, the more
significant cause of the shallower peak in previous IMF determinations for the inner ONC as opposed to the sharply peak IMF derived here
is likely just the inherent nature of photometric vs. spectroscopic mass
functions.  Determining masses for stars in a sample from temperatures derived
from spectral types necessarily discretizes the data.  
Conversely, photometric studies are by their nature 
continuous distributions, and
deriving masses from magnitudes and colors or luminosity
functions results in a smooth distribution of masses.  
We emphasize that while neither situation is ideal, mass functions 
derived for young objects from infrared photometry alone represents
only the most
statistically probable distribution of underlying masses.  
Spectroscopy is needed in 
order to derive cluster membership and masses for 
individual objects.

\subsection{Comparison to Other Star-Forming Regions}

Diagnostic studies of stellar populations in
different locations and at varying stages of evolution 
are needed to explore the possibility of a universal mass function.
While one might expect that the IMF should vary with 
star formation environment, we do not yet have enough evidence
to determine if such a variation exists.  
Aside from work already mentioned on the ONC, 
numerous studies have been carried out 
to characterize the low mass stellar and substellar 
mass functions of other young clusters in various 
environments.  
Because of the intrinsic faintness of these objects, most surveys are photometric.
Authors then use a combination of theoretical models and statistical
analysis to transform a cluster's color-magnitude diagram or 
luminosity function into an IMF which may or may not accurately represent the underlying cluster population (see Section 6.2).  

However, the substellar populations
of two other young
star-forming clusters have been studied spectroscopically
using techniques similar to those presented here.
Luhman (2000), Brice\~{n}o et al. (2002) and
Luhman et al. (2003a) surveyed the sparsely-populated 
Taurus star-forming region, 
and Luhman et al. (2003b) studied the rich cluster IC 348.
These clusters have ages similar to the ONC (1 and 2 Myr).
Therefore, if the IMF is universal, similar mass distributions
should be observed for all three clusters.  Contrary to this hypothesis,
Luhman et al. (2003b) discuss the very different shapes
of the substellar mass distributions in Taurus and IC 348.
The IMF for Taurus peaks around $\sim$0.8 M$_\odot$ and then declines
steadily to lower masses through the brown dwarf regime.
The IC 348 mass function rises to a peak around 0.15 M$_\odot$ 
and then falls off sharply and levels out for substellar objects.
While direct comparison
of the data requires caution given that different mass tracks were used
for the two studies (Luhman et al. (2003b) used Baraffe et al. (1998)
tracks whereas we have used DM97 tracks to infer masses),  
we find our that IMF for the ONC bears remarkable resemblance to
that presented for IC 348 in Luhman et al. (2003b) (see for comparison
Figure 12 in Luhman et al. 2003b).  Both IMFs peak at $\sim$0.15--0.2 M$_\odot$
and fall off rather abruptly at the substellar boundary.
The fact the IMFs derived for these two dense, young clusters 
bear such close resemblance to each other while exhibiting
distinguishable differences from the IMF determined for the much more sparsely-
populated Taurus cluster gives strong support to the argument
put forth by authors such as Luhman et al. (2003b) that 
the IMF is not universal, but may instead depend 
on star formation environment. 
Alternatively, Kroupa et al. (2003) argue through numerical simulation
that observed differences in the Taurus and ONC {\it stellar} mass
functions could be due to dynamical effects operating on
initially identical IMFs; however, their model does not reproduce the
observed differences in {\it substellar} mass functions without invoking
different initial conditions (eg., turbulence; Delgado-Donate et al. 2004).

Previous studies such as those mentioned above have found much
higher brown dwarf fractions for the ONC in comparison to other 
clusters such as Taurus and IC 348 (see Luhman et al. 2003b).  
Brice\~{n}o et al. (2002) defines the 
ratio of the numbers of substellar and stellar objects as:
\begin{displaymath}
R_{SS} = \frac{N(0.02 \leq M/M_\odot \leq 0.08)}{N(0.08 < M/M_\odot \leq 10)}.
\end{displaymath}
We have recomputed these numbers for Taurus and IC 348 using the DM97 models and find
values of $R_{SS}$ = 0.11 (Taurus) and $R_{SS}$ = 0.13 (IC 348).  These values
are very close to those found by Luhman et al. (2003b) using the Baraffe et al. (1998)
models: $R_{SS}$ = 0.14 (Taurus) and $R_{SS}$ = 0.12 (IC 348).  
In contrast, Luhman et al. (2000) find a value of $R_{SS}$ = 0.26 from primarily photometric work
on the inner ONC (using Baraffe et al. (1998) models).
Considering the spectroscopic IMF presented here (Figure 14), we find
a lower value of $R_{SS}$=0.20, indicating that although the ONC may have 
a higher brown dwarf fraction than Taurus or IC348, it is lower than previously
inferred from photometric studies.  


\section{Summary}

We have identified for the first time a large sample of spectroscopically confirmed
 young brown dwarfs within the inner regions of the ONC, including five newly 
classified M9 objects.  
From this data we have made
an HR diagram and possibly discovered a previously unknown population
of $\sim$10 Myr old low mass stars within the inner regions of the cluster.  We have examined this 
population in detail and determined that it is not an artifact 
arising from systematics in the spectral reduction/classification processes.
We have ruled out that this population consists primarily of ``contaminants''
from the surrounding OB1 association.  
If this population is real, we propose that it was not detected
in previous works because of its intrinsic faintness, coupled with extinction
effects.
Another possible scenerio is that these objects constitute a population 
of scattered light sources.  If this hypothesis is correct, these objects provide indirect 
observational support for recent theoretical arguments indicating higher disk occultaion fractions
for young substellar objects in comparison to higher mass CTTS (Walker et al. 2004).

From the HR diagram we have used DM97 tracks and isochrones to 
infer a mass for each object and constructed the first spectroscopic
substellar (0.6 $> M >$ 0.02 M$_\odot$)
IMF for the ONC.  Our mass function peaks at $\sim$0.2 M$_\odot$
consistent with previous IMF determinations (M02; HC00; Luhman et al. 2000),
however it drops off more sharply past $\sim$0.1 M$_\odot$ before roughly 
leveling off through the substellar regime.
We compare our mass function to those derived for Taurus and IC 348, also 
using spectroscopic data, and find remarkable agreement between 
mass distributions found in IC 348 and the ONC.  

\clearpage

\section{Acknowledgments}
The authors would like to thank Davy Kirkpatrick for useful comments
which improved the final paper, and Kevin Luhman and Lee Hartmann for their
insights and suggestions which helped in our analysis. 
We thank Alice Shapley and Dawn Erb for sharing their method
of subtracting sky lines from NIRSPEC images, as well as NIRSPEC
support astronomers Greg Wirth, Grant Hill and Paola Amico for their guidance during observing.
We are also appreciative of Michael Meyer for his participation in the CRSP data
acquisition and for assistance at the KPNO 4m telescope from Dick Joyce.
Finally, we thank LRIS observers N. Reid and B. Schaefer for obtaining several spectra for us.
C. L. S. acknowledges support from a National Science Foundation Graduate Research Fellowship. 

\clearpage
%

%
%
%

\clearpage
\input{tab1a.tex}

\clearpage
\input{tab1b.tex}


\clearpage
\input{tab2.tex}

\clearpage
\input{tab3.tex}

\clearpage

\title{\bf{\large{Erratum: ``The Spectroscopically Determined Substellar
Mass Function of the Orion Nebula Cluster'' (ApJ, 610, 1045 [2004])}}}

\author{Catherine L. Slesnick\altaffilmark{1}, Lynne
A. Hillenbrand\altaffilmark{1}, \& John M. Carpenter\altaffilmark{1}}

\affil{Dept.\ of Astronomy, MS105-24, California Institute of
Technology,Pasadena, CA 91125}

\email{cls@astro.caltech.edu, lah@astro.caltech.edu, jmc@astro.caltech.edu}

\altaffiltext{1}{Visiting astronomer, W. M. Keck Observatory, which
is operated as a scientific partnership among the California Institute of Technology, the university of California and the National Aeronautics and Space Administration.}

In Table 1 object HC-722 was incorrectly classified as an L0 dwarf.
The original classification was based on the apparent prominence
in the spectrum of the KI doublet ($\sim$7700 \AA). However, standard stars 
available to us at the time were of
lower resolution (by a factor of $\sim$2) compared to the program stars.
Narrow spectral features such as the KI doublet were thus 
blurred and appeared weaker.  We have since re-analyzed 
the spectrum of object HC-722 in comparison to higher resolution standards 
taken with LRIS (the same instrument 
used to obtain the spectrum of HC-722), and re-classified it as an 
M6.5 V dwarf object.  In table 4 the values for HC-722 should be replaced by:
log$T_{eff}$=3.429K, log$L$/L$_\odot$=-2.89, M=0.04 M$_\odot$.

Note that the object is still considered a dwarf, showing no signs of lower
gravity.  This, combined with the fact that we derive zero A$_V$ for HC-722,
implies it is likely a foreground field dwarf.
If we assume an absolute magnitude M$_K$=9.85 mag (Leggett 1992)
for an M6.5V star, we derive a distance modulus of 7.58 ($d\sim$330 pc).
Because HC-722 was previosly determined to have higher gravity than members
 of the ONC it was not included in 
the derivation or analysis of the cluster IMF.  As this ``high gravity'' 
assigmnment is not affected by the re-classification, the change 
does not alter or impact any of our original conclusions.

\end{document}

%% file: tab1a.tex


\begin{deluxetable}{cccccccccl}
\tabletypesize{\scriptsize}
\tablecolumns{10}
\tablenum{1a}
\tablecaption{Band Indices \& Spectral Types for the Inner 5.'1$\times$5.'1 of the ONC}
\tablehead{
  \colhead{ID\tablenotemark{a}} & 
  \colhead{$K$\tablenotemark{a}} &
  \multicolumn{2}{c}{$J$-Band} &
  \multicolumn{2}{c}{$K$-Band} &
  \multicolumn{2}{c}{Spectral Type} & 
  \colhead{Gravity\tablenotemark{d}} & 
  \colhead{Comments\tablenotemark{e}} \\ 
  \cline{3-4} 
  \cline{5-6}
  \cline{7-8} \\ 
  \colhead{} &
  \colhead{} &
  \colhead{H$_2$O-1} &
  \colhead{FeH} &
  \colhead{H$_2$O-2} &
  \colhead{H$_2$} &
  \colhead{Optical\tablenotemark{b}} &
  \colhead{IR\tablenotemark{c}} &
  \colhead{}  & 
  \colhead{}  
}
\startdata
    4 & 14.91 & 0.799 & 1.017 & \nodata & \nodata & M4.5 & M5.5  & low &  \\ 
   15 & 15.36 & 0.921 & 0.974 & \nodata & \nodata & M0-M1 & M3.5  & \nodata &  \\ 
   20 &  14.19 & 0.639 & 0.959 & 0.985 & -0.018 & \nodata & M8 &   low   & \\
   22 &  13.19 & \nodata & \nodata & 0.997 & -0.014 & \nodata &             M8 &   int   & possible NIR emission lines\\
   25 &  13.90 & \nodata & \nodata & 1.189 & -0.167 & M:\tablenotemark{f} &         M1 &   low/int   & \\
   27 & 16.30 & 0.801 & 0.962 & \nodata & \nodata & \nodata & M5  & low &  \\ 
   29 &  11.74 & 0.827 & 0.987 & 1.032 & -0.040 & \nodata &           M5 &   low   & \\
   30 & 15.05 & 0.877 & 1.020 & \nodata & \nodata & M0-3 & M2  & low & CaIIe, possible NIR emission lines \\ 
   35 & 11.11 &  \nodata &  \nodata &  \nodata &  \nodata & M3.5\tablenotemark{f}  & M7:\tablenotemark{g} &  \nodata & \\
   48 & 15.57 & 0.839 & 0.974 & \nodata & \nodata & \nodata & M4  & low &  \\ 
   51 &  11.63 & \nodata & \nodata & 1.017 & -0.013 & \nodata &           M5 &   int   & \\
   55 & 14.10 & 0.736 & 0.934 & \nodata & \nodata & \nodata & M8  & low & possible NIR emission lines \\ 
   59 & 16.29 & \nodata & \nodata & \nodata & \nodata & K8-M3 & \nodata & \nodata & \\ 
   62 &  15.34 & \nodata & \nodata & 0.879 &  0.039 & \nodata &          M9 &   low   & \\
   64 & 14.52 & \nodata & \nodata & \nodata & \nodata & M7-9 & \nodata & low & \\
   70 & 15.75 & 0.624 & 0.870 & \nodata & \nodata & \nodata & M9  & low &  \\ 
   90 &  13.01 & \nodata & \nodata & 0.953 &  0.030  & \nodata &             M7.5 &   int   & \\
   91 & 14.90 & 0.892 & 1.093 & \nodata & \nodata & \nodata & $<$K7\tablenotemark{h} & \nodata & possible NIR emission lines \\ 
   93 &  9.50 & 0.859 & 0.982 & \nodata & \nodata & M3e\tablenotemark{f} & M3  & high & possible NIR emission lines \\ 
  111 & 15.38 & 0.704 & 0.943 & \nodata & \nodata & \nodata & M9  & low &  \\ 
  114 & 13.94 & \nodata & \nodata & \nodata & \nodata & M7 & \nodata & low &  \\ 
  123 &  14.61 & \nodata & \nodata & 0.977 &  0.031 & M0-5 &             M7.5 &   low & \\  
  127 & 14.55 & 0.960 & 1.021 & \nodata & \nodata & \nodata & M0  & int &  \\ 
  143 &  15.00 & \nodata & \nodata & 1.152 & -0.377 & $<$M2: &           M2.5 &   low   & \\
  162 &  12.48 & \nodata & \nodata & 0.979 & -0.035 & \nodata &             M5.5 &   low   & \\
  167 &  15.02 & \nodata & \nodata & 0.942 & -0.007 & M6-8 &  M7.5 &   low   & \\
  200 &  15.49 & \nodata & \nodata & 1.145 & -0.159 & \nodata &             M3 & \nodata   & \\
  210 & 14.68 & 0.732 & 0.996 & \nodata & \nodata & $>$M6: & M7  & low &  CaIIe\\ 
  212 & 14.22 & 0.683 & 0.904 & \nodata & \nodata & \nodata & M9  & low &  \\ 
  221 &  14.46 & \nodata & \nodata & 0.949 & -0.038 & \nodata &             M7.5 &   low   & \\
  227 & 10.96 & 0.892 & 1.105 & \nodata & \nodata & M1 & $<$K7\tablenotemark{h}  & \nodata &  \\ 
  237 & 13.44 & \nodata & \nodata & \nodata & \nodata & M2 & \nodata & low & CaIIe \\
  264 & 12.41  & \nodata &  \nodata &  \nodata &  \nodata &  M6\tablenotemark{f} &  M6\tablenotemark{g} & \nodata & \\ 
  288 & 10.62 & 0.921 & 0.982 & \nodata & \nodata & K8\tablenotemark{f} & M1  & high & possible NIR emission lines \\ 
  290 &  13.47 & \nodata & \nodata & 0.989 & -0.013 & \nodata &         M5.5: &  \nodata   & \\
  293 &  9.14 & 0.700 & 1.283 & \nodata & \nodata & B8 & $<$K7\tablenotemark{h}  & low &  \\
  316 &  15.18 & \nodata &  \nodata &  \nodata &  \nodata & M3.5-5 & \nodata & low & CaIIe  \\ 
  346 &  10.48 & \nodata & \nodata & 1.120 & -0.106 & \nodata &            M2 &  low/int   & \\
  355 &  12.35 & \nodata & \nodata & 1.054 & -0.067 & \nodata &            M5 &   int   & possible NIR emission lines  \\
  365 &  15.40 & \nodata & \nodata & 0.918 &  0.004 & \nodata &             M7: &     \nodata   & \\
  366 &  12.68 & \nodata & \nodata & 0.926 &  0.024 & \nodata &          M7.5 &   low   & possible NIR emission lines  \\
  368 &  11.07 & \nodata & \nodata & 1.014 & -0.025 & \nodata &         M5 &   int   & possible H$_2$ and HeI P cygni profile \\
  372 & 14.48 & 0.677 & 0.958 & \nodata & \nodata & \nodata & M9  & low &  \\ 
  381 &  11.06 & \nodata & \nodata & 1.129 & -0.112 & M2\tablenotemark{f} &         M1.5-4 &   int   & \\
  383 &  14.36 & 0.922 & 0.869 & 1.066 & -0.121 & \nodata &         M4 &   low   & possible NIR emission lines\\
  395 &   9.83 & \nodata & \nodata & 1.016 & -0.040 & \nodata &        M2: &     \nodata   & \\
  396 &  11.93 & \nodata & \nodata & 1.006 & -0.026 & M:\tablenotemark{f} &            M4-8 &     int   & possible NIR emission lines\\
  400 & 15.16 & 0.694 & 0.913 & \nodata & \nodata & M7-8 & M9  & low &  \\ 
  403 &  15.15 & \nodata & \nodata & 0.930 &  0.005 & \nodata &            M7 &   low   & possible NIR emission lines\\
  409 & 14.39 & \nodata &  \nodata &  \nodata &  \nodata & M0 & \nodata & \nodata & \nodata  \\
  429 &  15.11 & \nodata & \nodata & 0.938 &  0.031 & M7-9 &          M7.5 &   low   & \\
  433 & 15.58 & 0.695 & 0.994 & \nodata & \nodata & \nodata & M8  & low &  \\ 
  434 &  10.04 & \nodata & \nodata & 0.949 &  0.022 & \nodata &       M2 & \nodata  & possible HeI P cygni profile\\
  454 &  12.06 & \nodata & \nodata & 1.055 & -0.068 & \nodata &           M2 &  high   & \\
  455 &  15.84 & \nodata & \nodata & 0.995 & -0.066 & \nodata &        M2-6 &     \nodata   & \\
  459 &  12.34 & \nodata & \nodata & 1.072 & -0.051 & \nodata &           M6 &  int   & \\
  467 &  11.17 & \nodata & \nodata & 1.056 & -0.050 & M2-4\tablenotemark{f} &             M6 &  high   & \\
  469 & 14.74 & \nodata &  \nodata &  \nodata &  \nodata & M &  \nodata & \nodata & \\
  509 &  15.33 & 0.900 & 0.988 & 0.986 & -0.126 & M2-5 &           M2-7 &  \nodata  & \\
  515 &  15.88 & \nodata & \nodata & 0.804 &  0.088 & \nodata &            M7: &   low   & \\
  529 &  16.17 & \nodata & \nodata & 0.775 &  0.151 & \nodata &          M8 &   low   & \\
  543 & 10.27 & 0.931 & 0.977 & \nodata & \nodata & M2.5\tablenotemark{f} & M1  & int/high &  possible NIR emission lines \\ 
  553 &  12.31 & \nodata & \nodata & 1.043 & -0.035 & \nodata &             M5 &   low/int   & possible NIR emission lines\\
  555 &  11.58 & \nodata & \nodata & 1.090 & -0.049 & \nodata &           M6 &  \nodata   & \\
  559 &  14.40 & \nodata & \nodata & 0.958 & -0.017 & \nodata &            M8 &   low   & possible NIR emission lines\\
  565 &  15.00 & \nodata & \nodata & 0.930 &  0.048 & \nodata &           M8 &   low   & \\
  568 &  15.00 & \nodata & \nodata & 0.808 &  0.172 & \nodata &             $<$K7\tablenotemark{h}: &   \nodata   & \\
  594 & 14.58 & 0.776 & 1.039 & \nodata & \nodata & \nodata & M7.5  & low &  \\ 
  600 &  12.25 & \nodata & \nodata & 0.979 &  0.056 & \nodata &             M5 &   low   & \\
  684 &  14.66 & \nodata & \nodata & 1.094 & -0.081 & \nodata &          M0-4 &     \nodata   & \\
  687 &  15.28 & \nodata & \nodata & 1.110 & -0.118 & \nodata &          M0-4 &  \nodata   & \\
  708 &  15.04 & \nodata & \nodata & 0.905 &  0.098 & \nodata &             M4 &   low   & \\
  709 &  16.06 & \nodata & \nodata & 0.854 &  0.041 & \nodata &            M5 &   low   & \\
  721 &  15.18 & \nodata & \nodata & 0.995 & -0.102 & \nodata &           M3.5: &   \nodata   & \\
  722 & 17.43 &  \nodata &  \nodata &  \nodata &  \nodata & L0 &  \nodata & high & \\
  723 & 16.07 &  \nodata &  \nodata &  \nodata &  \nodata & M3 &  \nodata & low & \\
  724 &  16.51 & \nodata & \nodata & 1.218 & -0.044 & \nodata &         M6: &   \nodata   & \\
  725 &  16.07 & \nodata & \nodata & 0.916 &  0.041 & \nodata &            M7: &   \nodata   & \\
  728 & 16.34 & 0.772 & 1.074 & \nodata & \nodata & \nodata & M5.5  & low &  \\ 
  729 &  16.02 & \nodata & \nodata & 0.861 &  0.032 & \nodata &           M7 &   low   & possible NIR emission lines\\
  730 &  15.15 & \nodata & \nodata & 1.111 & -0.114 & \nodata &           M4 &   low   & possible NIR emission lines\\
  731 &  15.79 & \nodata & \nodata & 1.061 & -0.125 & \nodata &             K7: &     \nodata  &  \\
  732 &  16.80 & \nodata & \nodata & 1.029 &  0.046 & \nodata &             M2.5: &     \nodata   & \\
  743 &  16.89 & \nodata & \nodata & 0.997 & -0.055 & \nodata &       M6: &     \nodata   & \\
  749 &  17.40 & \nodata & \nodata & 1.001 &  0.075 & \nodata &        M8 &     \nodata   & \\
  764 &  14.62 & \nodata & \nodata & 0.974 & -0.194 & K8-M3 &   M7.5: &     \nodata   & \\
\enddata

\tablewidth{40pc}
\tablenotetext{a}{IDs and $K$ magnitudes from Hillenbrand \& Carpenter, 2000.}
\tablenotetext{b}{Optical spectral types are from LRIS data presented here, unless otherwise noted.}
\tablenotetext{c}{Uncertainties in infrared spectral types are $\pm$1.5 sub-class unless otherwise noted.
A ":" indicates the spectrum had lower S/N and the classification is less certain.}
\tablenotetext{d}{For $J$-band spectra, an object was given a gravity classification of ''low" if it had no detectable
atomic absorption lines and ''high" if it had strong lines, similar to those seen in spectra of dwarf standards
of same spectral types.
A classification of ``int" indicates absorption lines were present but not as strong as those in dwarf stars at the 
same temperatures.  For $K$-band spectra gravity classifications were determined based
on the relative atomic line ratios.  A "\nodata" indicates the S/N of the spectra was not sufficient to determine if absorption lines were present.}
\tablenotetext{e}{A comment of ``possible NIR emission lines'' indicates that residual
emission lines were present after background sky subtraction which may 
originate from the star rather than the nebula.  However, due to the very
high background of nebular emission, we cannot say with certainty that this is the case.  
This comment primarily refers to H$_2$ seen in emission.  P cygni profile
labels are given as uncertainties for the same reason.}
\tablenotetext{f}{Optical spectral type from Hillenbrand 1997.}
\tablenotetext{g}{Classification from CRSP spectrum.}
\tablenotetext{h}{We did not attempt to classify objects earlier than K7.}
\end{deluxetable}


%% file: tab1b.tex


\begin{deluxetable}{ccccl}
\tabletypesize{\scriptsize}
\tablecolumns{4}
\tablewidth{0pc}
\tablenum{1b}
\tablecaption{Spectral Types for the Outer\tablenotemark{a}$\;$ ONC}
\tablehead{
  \colhead{} &
  \colhead{} &
  \multicolumn{2}{c}{Spectral Type} \\
  \cline{3-4}  \\
  \colhead{ID\tablenotemark{b}} & 
  \colhead{$K$\tablenotemark{b}} &
  \colhead{Optical\tablenotemark{c}} &
  \colhead{IR\tablenotemark{d}} &
}
\startdata
1      & 12.49  & M5.5\tablenotemark{e} & M6 \\  
205    & 11.68  & M6\tablenotemark{e} & M7 \\  
615    & 12.32  & M6\tablenotemark{e} & M7 \\  
649    & 12.13  & M6\tablenotemark{e} & M8 \\  
689    & 12.75  & M5.5\tablenotemark{e} & M6 \\  
711    & \nodata  & M6\tablenotemark{e} & M7: \\  
806    & 12.32  & M5.5 & M6.5\\  
859    & 11.75  & M5\tablenotemark{e} & M8\\  
974    & 11.38  & M5.5 & M6-8\\  
1036   & 12.38  & M6\tablenotemark{e} & M6 \\  
3017   & 12.62  & M4.5\tablenotemark{e} & M2-5 \\   
3039   & 11.76  & M5.5 & M7 \\  
3046   & 11.79  & M5.5 & M5 \\  
5100   & 11.81  & M5.5e\tablenotemark{e} & M6 \\ 
\enddata

\tablewidth{40pc}

\tablenotetext{a}{Outer in this context refers to beyond 
the 5.'1$\times$5.'1 NIRC field centered on $\theta ^1$C.}
\tablenotetext{b}{IDs and $K$ magnitudes from Hillenbrand et al., 1998.  See references therein.}
\tablenotetext{c}{Optical spectral types are from LRIS data, unless otherwise noted.}
\tablenotetext{d}{Infrared spectral type uncertainties are $\pm$2 sub-class unless otherwise noted.  A ":" indicates the spectrum had lower S/N and the classification is less certain.}
\tablenotetext{e}{Optical spectral types come from Hillenbrand 1997.}

\end{deluxetable}


%% file: tab2.tex


\begin{deluxetable}{lll|c}
\rotate
\tabletypesize{\scriptsize}
\tablecolumns{3}
\tablewidth{0pc}                                  
\tablenum{2}                                    
\tablecaption{Empirical Relations}
\label{tbl:tab2}
\tablehead{
  \colhead{Fit} &
  \colhead{Error} &
  \colhead{Range} &
  \colhead{Comment} 
}
\startdata
 Sp type $\;=\;$ (33.71 $\pm$0.79) - (35.35 $\pm$1.06)*(H$_2$O-1)  &  $\sigma$ = $\pm$1.2 & M0-T0 &
In the spectral type relations for band index fits M0 is \\
 Sp type $\;=\;$ (36.31 $\pm$1.46) - (31.59 $\pm$1.59)*(FeH)  &  $\sigma$ = $\pm$0.66 & M3-L3 & respresented by 0, M5 by 5, and L0 by 10.  All Sp type \\ 
 Sp type $\;=\;$ (34.13 $\pm$1.19) - (27.10 $\pm$1.20)*(H$_2$O-2)  &  $\sigma$ = $\pm$0.53  & M2-L3 & equations assume dereddened band index measurements.\\ 
\\
T$_{eff}$ $\;=\;$ -224536 + 111513*(Sp type) - 17907.9*(Sp type)$^2$ + 943.944*(Sp type)$^3$ &  $\sigma$ = $\pm$33.9 & K7-L0 & In the spectral type relations for HR diagram transforms\\
$(H-K)_o$ $\;=\;$ 51.4776 - 23.9402*(Sp type) + 3.67070*(Sp type)$^2$ - 0.184388*(Sp type)$^3$ &  $\sigma$ = $\pm$0.033 & K7-L0 &  M0 is represented by 6.0, M5 by 6.5, and L0 by 7.0. \\
BC$_K$ $\;=\;$ 57.1637 - 27.7257*(Sp type) + 4.54411*(Sp type)$^2$ - 0.240687*(Sp type)$^3$ &  $\sigma$ = $\pm$0.037 & K7-L0 & \\

\enddata
\end{deluxetable}

%% file: tab3.tex
\begin{deluxetable}{ccccc}
\tabletypesize{\scriptsize}
\tablecolumns{5}
\tablewidth{0pc}
\tablenum{3}
\tablecaption{Derived Quantities for Stars in the HR Diagram}
\tablehead{
  \colhead{ID\tablenotemark} & 
  \colhead{A$_V$ [mag]}&
  \colhead{log $T_{eff}$ [K]} &
  \colhead{log $L$/L$_\odot$} &
  \colhead{$M$/M$_\odot$} \\
}
\startdata
  4 & 4.67 & 3.453 & -1.686 & 0.08 \\ 
 15 & 8.47 & 3.502 & -1.662 & 0.19 \\ 
 20 & 3.77 & 3.397 & -1.507 & 0.04 \\ 
 22 & 1.78 & 3.397 & -1.178 & 0.05 \\ 
 25 & 6.00 & 3.556 & -1.080 & 0.42 \\ 
 27 & 9.00 & 3.465 & -2.072 & 0.08 \\ 
 29 & 4.77 & 3.465 & -0.400 & 0.11 \\ 
 30 & 2.18 & 3.535 & -1.713 & 0.31 \\ 
 35 & 1.94 & 3.418 & -0.315 & 0.10 \\ 
 48 & 5.37 & 3.490 & -1.876 & 0.15 \\ 
 51 & 2.26 & 3.465 & -0.447 & 0.11 \\ 
 55 & 1.90 & 3.397 & -1.540 & 0.03 \\ 
 59 & 0.02 & 3.556 & -2.251 & 0.34 \\ 
 62 & 5.85 & 3.380 & -1.920 & 0.02 \\ 
 64 & 4.46 & 3.397 & -1.617 & 0.03 \\
 70 & 9.48 & 3.380 & -1.953 & 0.02 \\ 
 90 & 6.77 & 3.407 & -0.915 & 0.07 \\ 
 93 & 8.24 & 3.513 & 0.692 & 0.18 \\ 
111 & 8.61 & 3.380 & -1.837 & 0.03 \\ 
114 & 1.44 & 3.418 & -1.464 & 0.05 \\ 
123 & 7.89 & 3.407 & -1.515 & 0.04 \\ 
127 & 12.51 & 3.574 & -1.072 & 0.58 \\ 
143 & 6.31 & 3.525 & -1.560 & 0.27 \\ 
162 & 9.17 & 3.453 & -0.553 & 0.11 \\ 
167 & 9.19 & 3.407 & -1.631 & 0.03 \\ 
200 & 4.58 & 3.513 & -1.838 & 0.22 \\ 
210 & 4.81 & 3.418 & -1.636 & 0.04 \\ 
212 & 3.41 & 3.380 & -1.559 & 0.03 \\ 
221 & 4.43 & 3.407 & -1.577 & 0.04 \\ 
237 & 1.44 & 3.535 & -1.097 & 0.29 \\ 
264 & 1.67 & 3.441 & -0.810 & 0.10 \\ 
288 & 4.08 & 3.556 & 0.161 & 0.23 \\ 
290 & 6.64 & 3.453 & -1.039 & 0.11 \\ 
316 & 10.89 & 3.483 & -1.529 & 0.15 \\ 
346 & 5.81 & 3.535 & 0.247 & 0.18 \\ 
355 & 3.32 & 3.465 & -0.695 & 0.12 \\ 
365 & 8.23 & 3.418 & -1.804 & 0.04 \\ 
366 & 10.25 & 3.407 & -0.657 & 0.08 \\ 
368 & 8.77 & 3.465 & 0.014 & 0.10 \\ 
372 & 2.04 & 3.380 & -1.711 & 0.03 \\ 
381 & 3.93 & 3.519 & -0.079 & 0.17 \\ 
383 & 9.35 & 3.490 & -1.247 & 0.16 \\ 
395 & 12.41 & 3.535 & 0.744 & 0.20 \\ 
396 & 13.53 & 3.441 & -0.193 & 0.10 \\ 
400 & 2.16 & 3.380 & -1.978 & 0.02 \\ 
403 & 6.78 & 3.418 & -1.754 & 0.04 \\ 
409 & 12.51 & 3.574 & -1.010 & 0.59 \\ 
429 & 5.53 & 3.407 & -1.798 & 0.03 \\ 
433 & 10.44 & 3.397 & -1.823 & 0.02 \\ 
434\tablenotemark{a} & 21.41 & 3.535 & 0.983 & 0.20 \\ 
454 & 7.85 & 3.535 & -0.314 & 0.23 \\ 
455 & 8.21 & 3.490 & -1.881 & 0.15 \\ 
459 & 1.59 & 3.441 & -0.786 & 0.10 \\ 
467 & 2.55 & 3.441 & -0.284 & 0.10 \\ 
509 & 4.86 & 3.477 & -1.813 & 0.12 \\ 
515 & 23.29 & 3.418 & -1.452 & 0.05 \\ 
529 & 25.18 & 3.397 & -1.530 & 0.03 \\ 
543 & 4.00 & 3.556 & 0.299 & 0.21 \\ 
553 & 5.77 & 3.465 & -0.592 & 0.11 \\ 
555 & 1.56 & 3.441 & -0.482 & 0.10 \\ 
559 & 0.00 & 3.397 & -1.729 & 0.03 \\ 
565 & 5.60 & 3.397 & -1.764 & 0.02 \\ 
594 & 2.20 & 3.407 & -1.707 & 0.03 \\ 
600 & 14.26 & 3.465 & -0.263 & 0.11 \\ 
684 & 7.93 & 3.535 & -1.349 & 0.31 \\ 
687 & 6.68 & 3.535 & -1.641 & 0.31 \\ 
708 & 28.72 & 3.490 & -0.822 & 0.15 \\ 
709 & 2.11 & 3.465 & -2.223 & 0.07 \\ 
721 & 9.38 & 3.502 & -1.558 & 0.19 \\ 
722 & 0.00 & 3.369 & -2.989 & 0.02 \\ 
723 & 3.55 & 3.513 & -2.106 & 0.21 \\ 
724 & 0.00 & 3.441 & -2.511 & 0.03 \\ 
725 & 9.40 & 3.418 & -2.030 & 0.03 \\ 
728 & 5.91 & 3.453 & -2.213 & 0.05 \\ 
729 & 14.73 & 3.418 & -1.818 & 0.04 \\ 
730 & 15.69 & 3.490 & -1.335 & 0.17 \\ 
731 & 19.53 & 3.610 & -1.228 & 0.56 \\ 
732 & 0.00 & 3.525 & -2.508 & 0.23 \\ 
743 & 5.61 & 3.441 & -2.462 & 0.03 \\ 
749 & 3.10 & 3.397 & -2.815 & 0.02 \\ 
764 & 7.06 & 3.407 & -1.548 & 0.04 \\ 
\enddata

\tablewidth{40pc}
\tablenotetext{a}{Based on visual inspection of the spectra of HC 434, we find 
the extinction value derived for this source using the
intrinsic colors of a spectral type M2 star 
to be an over-estimate.  Instead we use the extinction 
derived by dereddening its ($J-H$),($H-K$) colors back to the CTTS locus (see section
3.4) to determine the 
derived quantities listed in the table as well to place HC 434 on the 
HR diagram (Fig.~\ref{fig:hrd}).}
\end{deluxetable}